%% file: main.tex
\documentclass[sigconf,authorversion,nonacm]{acmart}

\AtBeginDocument{%
  }

\usepackage{multirow}
\usepackage{graphicx}
\usepackage{microtype}

\begin{document}

\title[AdaptType]{\textit{AdaptType}: Accessible and Customizable Multimodal Text Entry in Virtual Reality for People with Upper-Body Motor Impairments}

\author{Junhan Kong}
\affiliation{%
  \institution{The Information School\\University of Washington}
  \city{Seattle}
  \state{Washington}
  \country{USA}
}
\email{junhank@uw.edu}

\author{Mingyuan Zhong}
\affiliation{%
  \institution{Computer Science \& Engineering\\University of Washington}
  \city{Seattle}
  \state{Washington}
  \country{USA}
}
\email{myzhong@cs.washington.edu}

\author{Sanjana Pavani}
\affiliation{%
  \institution{The Information School\\University of Washington}
  \city{Seattle}
  \state{Washington}
  \country{USA}
}
\email{sanjjjp@uw.edu}

\author{Jacob O. Wobbrock}
\affiliation{%
  \institution{The Information School\\University of Washington}
  \city{Seattle}
  \state{Washington}
  \country{USA}
}
\email{wobbrock@uw.edu}

\renewcommand{\shortauthors}{Kong et al.}

\begin{abstract}
Text entry in virtual reality (VR) is cumbersome, especially for people with upper-body motor impairments. Existing techniques often demand both fine and gross motor control, yet users' motor abilities vary widely. To address this range of needs, we present \textit{AdaptType}, a multimodal VR text entry system that enables users to configure input techniques that best suit their abilities, and scaffolds customization through ability-based recommendations. Through a study with 10 participants, we found that customized multimodal techniques improved text entry speed compared to gaze- or hand-only techniques, and complemented voice for error correction, decreasing failure rate from 15.1\% to 2.0\%. We found that \textit{AdaptType} yielded high text entry speed (36.26 WPM) and accuracy (0.36\% error rate), with low task load and high usability ratings. We also found ability-based recommendations effectively scaffolded customization. Our work contributes more accessible and personalized VR text entry solutions for people with upper-body motor impairments.
\end{abstract}


\begin{teaserfigure}
    \centering
    \includegraphics[width=0.95\textwidth]{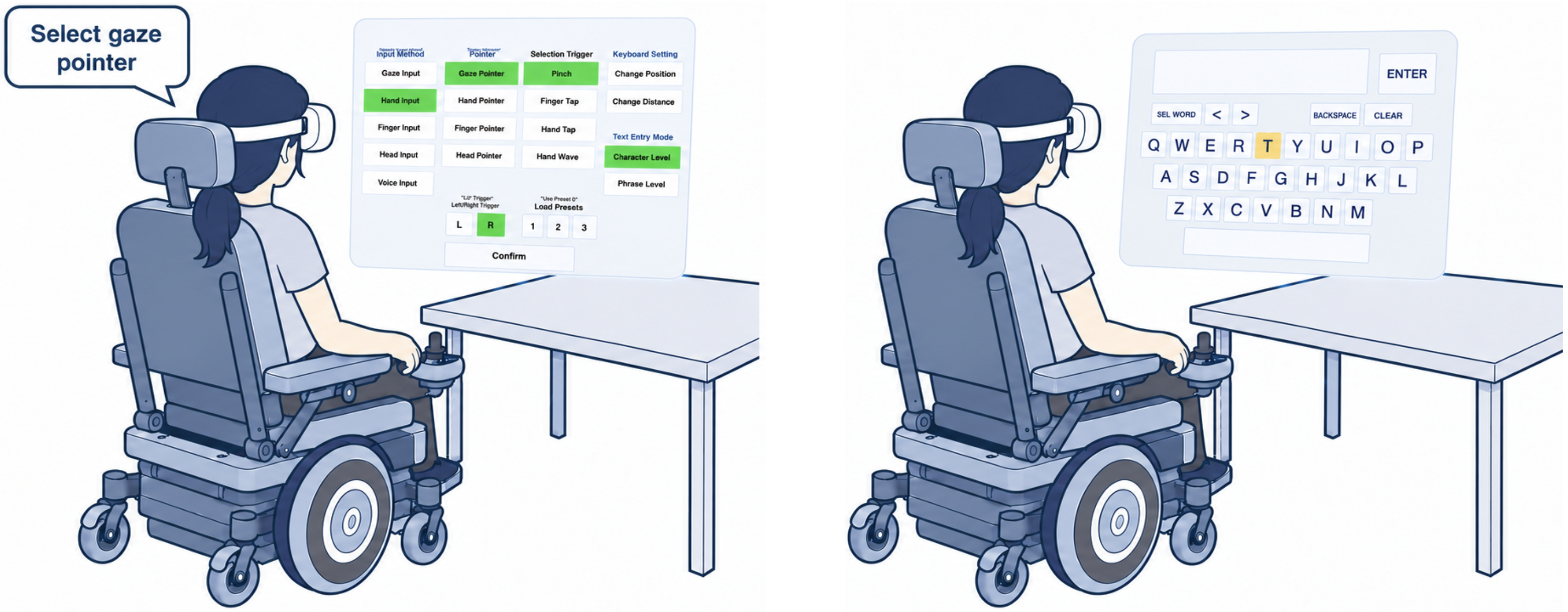}
    \caption{\textit{AdaptType} enables users with upper-body motor impairments to select VR text entry methods that best suit their abilities by offering conversational configurations of input modalities, pointer and selection triggers, keyboard settings, and more, scaffolding these end-user customizations through ability-based recommendations. \textit{AdaptType} makes VR text entry more accessible for users with upper-body motor impairments, exhibiting high text entry speed and accuracy, low task load, and high usability.}
    \Description{Two illustrations demonstrate AdaptType in use by a wheelchair user wearing a VR headset. In the left illustration, the user customizes the text entry method through a settings interface, with a voice command selecting a gaze pointer. In the right illustration, the user enters text on a virtual QWERTY keyboard using the resulting customized configuration.}
\end{teaserfigure}

\maketitle

\input{01-introduction}
\input{02-related-work}
\input{03-co-design}

\input{04-system-design}
\input{05-method}
\input{06-results}
\input{07-discussion}
\input{08-conclusion}

\begin{acks}
The authors thank the Here and Now Project, Sue Spaulding, Kathleen Quin Voss, Momona Yamagami, and Qingyang Wan. This work was supported in part by the University of Washington Center for Research and Education on Accessible Technology and Experiences (CREATE). Any opinions, findings, conclusions or recommendations expressed in our work are those of the authors and do not necessarily reflect those of any supporter.
\end{acks}

\bibliographystyle{ACM-Reference-Format}
\bibliography{reference}

\end{document}

%% file: 01-introduction.tex
\section{Introduction}

Despite the rapid development of virtual reality (VR) devices and sensing capabilities, text entry in VR systems remains cumbersome, especially for people with upper-body motor impairments. People with upper-body motor impairments often face significant barriers while interacting with VR systems~\cite{yamagami2023people, franz2019perception, wentzel2022understanding}, as most existing input techniques require both fine and gross motor control. 
Unlike text entry using physical keyboards or touchscreen devices, VR text entry often relies on mid-air pointing or gestures that require higher levels of dexterity~\cite{poupyrev1998egocentric, dube2019text}. These mid-air actions offer no physical support for the user to rest their hand, wrist, or arm (the ``gorilla arm problem''~\cite{hansberger2017dispelling, palmeira2023quantifying, jang2017modeling}), and thus may cause severe fatigue and discomfort, especially for people with upper-body motor impairments. 

Various techniques have been developed to support VR text entry, for example, by leveraging physical keyboards~\cite{grubert2018standard}, gaze~\cite{rajanna2018gaze, majaranta2002twenty}, or voice~\cite{pick2016swifter, ruan2018speech}. However, these techniques each have their own issues for people with motor impairments. Hand-based keyboard interactions may exclude users with limited dexterity; gaze input requires no motor function but can be relatively slow~\cite{rajanna2018gaze} and sometimes fatiguing~\cite{vasiljevas2016modelling}, and may interfere with other tasks such as browsing~\cite{maglio2000suitor}; voice dictation is often fast and easy to use, yet can be unreliable in noisy environments~\cite{derby2020we, sinlapanuntakul2022effects} or for users with impaired speech~\cite{findlater2010beyond, young2010difficulties}. Prior work also found that single-modality techniques are often insufficient to accommodate the diverse range of needs for people with upper-body motor impairments, and that one modality that works well for one person may be unusable for another~\cite{li2022freedom, findlater2010beyond}. 

Importantly, motor impairments are \textbf{\textit{heterogeneous}}: some people retain strong gross motor abilities but have limited finger dexterity, while others may experience the reverse~\cite{yamagami2023people}. This diverse range of needs suggests that there is unlikely a \textit{one-size-fits-all} solution for accessible VR text entry~\cite{harper2007there, shneiderman2003universal, stephanidis2001adaptive}. 

To address this wide range of user needs, we developed \textit{AdaptType}, a highly customizable multimodal VR text entry system that enables users to configure input techniques to suit their own abilities. The system implements a comprehensive end-user customization framework that includes five categories of customizable settings: (1) input modalities, (2) auxiliary pointers, (3) selection triggers, (4) keyboard interface settings, and (5) modes of text entry. Beyond selecting from a pre-defined set of pointing and triggering gestures, \textit{AdaptType} also allows users to specify customized gestures that suit their fine or gross motor abilities. To help users navigate the large number of options and reduce cognitive burden in customization, the system leverages large-language models (LLMs) to provide ability-based recommendations to scaffold the customization process. Based on these recommendations, the system allows users to further adjust customization settings through conversational interactions. 

Over multiple preliminary sessions with three people with upper-body motor impairments, we co-designed the \textit{AdaptType} system, beginning with an initial prototype we developed and using participants' feedback and insights to guide iterative refinements toward the final working system.
We then evaluated our final \textit{AdaptType} system through a lab study with 10 participants with a range of upper-body motor disabilities. During the study sessions, participants completed text entry tasks using their customized technique, as well as using hand-, gaze-, and voice-only techniques; participants then iterated on their customization choices based on the experience. 

Specifically, our evaluation answered two main research questions: 
\begin{itemize}
    \item \textbf{RQ1.} Do customized multimodal text entry methods make VR text entry more accessible for people with upper-body motor impairments? If so, in which respects?
    \item \textbf{RQ2.} What approaches can effectively support customization of VR text entry for people with upper-body motor impairments? 
\end{itemize}

Our study results showed that customized multimodal techniques improved text entry speed compared to gaze- or hand-only techniques, and complemented voice
for error correction, decreasing failure rate from 15.1\% to 2.0\%. We found that \textit{AdaptType} yielded high text entry speed (36.26 WPM) and accuracy (0.36\% error rate), with low task load and high usability ratings. Additionally, we found that \textit{AdaptType}'s ability-based recommendations effectively scaffolded selection of text entry methods that matched participants' motor abilities, and that iterative customization showed trends of improvement in text entry performance, task load, and usability. 

Note that the goal of our study was \textit{not} to provide a comprehensive comparison of all VR text entry modalities; nor was it to show that customized techniques \textit{outperform} every baseline in every respect. Rather, our study demonstrates how \textit{AdaptType} makes VR text entry accessible by \textit{accommodating} a range of motor abilities and \textit{enabling} selection of techniques that best match each user's abilities. Our findings also shed light on design implications for accessible personalized text entry, suggesting that text entry interfaces should offer ability-based recommendations to scaffold customization, and should guide users to iteratively refine their customization choices over time. 

The main contributions of this paper are:

\begin{itemize}
    \item A framework for end-user customization that supports accessible and multimodal text entry for people with upper-body motor impairments;
    \item The \textit{AdaptType} system, a mixed-initiative~\cite{horvitz1999principles} and ability-based~\cite{wobbrock2011ability, wobbrock2018ability} VR text entry system that offers a rich set of customization options and provides recommendations to scaffold customizations based on users' abilities; 
    \item Empirical findings on customization choices, input behavior, text entry performances, and preferences with different modalities in VR text entry for people with upper-body motor impairments. 
\end{itemize}

%% file: 02-related-work.tex
\section{Related Work}

\subsection{Ability-Based Design}

Our work builds upon the philosophy of Ability-Based Design (ABD) introduced by Wobbrock et al.~\cite{wobbrock2011ability, wobbrock2018ability}. ABD is a design approach that centers the abilities of users throughout the design process, with the goal of creating systems that ``leverage the full range of human potential''~\cite{wobbrock2011ability}. In an ability-based system, the responsibility for adaptation lies with the system. It is the system that adjusts to the user's specific abilities, rather than requiring the user to conform to the system's constraints. Systems might also be highly configurable by end-users so as to better match their abilities. 

A number of prior systems have been informed by ABD principles, from automatic user interface generation~\cite{gajos2007automatically, gajos2008improving, gajos2010automatically} to enhancing outdoor play experiences for children~\cite{alsaleem2020applying}. Recent systems adopt signals such as gaze, motion, and touch patterns to support adaptation, for example, characterizing specific fine motor challenges through touch behavior~\cite{kong2022quantifying}, improving touch accuracy for people with motor impairments~\cite{mott2016smart, mott2019cluster}, assisting with reading resumption~\cite{mariakakis2015switchback}, and supporting adaptive app creation~\cite{kong2024ability}. Within the domain of text entry, prior work proposed techniques for dynamically adapting keyboards based on detected walking\cite{goel2012walktype} or hand posture~\cite{goel2013contexttype} for improved accuracy, and ability-based optimization of keyboard layouts based on user movement patterns~\cite{mitchell2022ability} and for older adults ~\cite{sarcar2016towards, sarcar2017ability, sarcar2018ability}. 

While much of the existing work in ABD emphasizes \textit{adaptivity}---that is, systems that automatically adapt based on sensed or inferred user abilities---there is a limited body of work focused on \textit{adaptability}, where users themselves actively configure or customize the system to suit their unique abilities. Our work offers a new perspective by supporting adaptability in the context of VR text entry, allowing users to personalize their interactive experiences based on their specific motor abilities.

\subsection{Virtual Reality Accessibility for Upper-Body Motor Impairments}
Current VR technologies present substantial barriers to people with disabilities~\cite{gerling2021critical}. Mott et al.~\cite{mott2020assuming} identified seven categories of access barriers in VR and highlighted the need for design for \textit{customization} and \textit{diversity}, and through an \textit{ability-based} approach. Gerling et al.~\cite{gerling2025equitable} also emphasized designing for \textit{experience} for equitable access to VR. A review by Dudley et al.~\cite{dudley2023inclusive} on VR/AR accessibility further highlighted the need of accommodating a combination of impairments rather than optimizing for single conditions.

Prior work has explored accessible VR interaction through scene-viewing and point-of-interest techniques~\cite{franz2019perception, franz2021nearmi}, mappings between bimanual and unimanual interactions~\cite{yamagami2022two}, and multimodal input and geometric motion mapping~\cite{wentzel2022understanding, wentzel2025motionblocks}, and highlighted accessibility as a key design dimension in locomotion techniques~\cite{di2021locomotion}. In particular, Di Luca et al.~\cite{di2021locomotion} found that locomotion techniques may be more accessible when they are abstract, and less accessible when they resemble real-world movement.
Most recently, Pococke et al.~\cite{pococke2025freehand} showed that default freehand gestures, including pinch and grasp, cause accessibility barriers because they rely heavily on the index finger. Franz et al.~\cite{franz2023comparing} also suggested having a range of accessible techniques to choose from based on user preferences. These findings directly motivated our decision to make the selection trigger configurable rather than defaulting to pinch.

\subsection{User-Defined Gestures}

A complementary line of work approached accessible input through user-defined gestures, in both traditional 2D GUIs and 3D interfaces. Following Wobbrock et al.'s work on eliciting user-defined gestures~\cite{wobbrock2005maximizing, wobbrock2009user} and the \$1 unistroke recognizer~\cite{wobbrock2007gestures}, subsequent recognizers have supported multistroke~\cite{anthony2010lightweight, anthony2012n} and point-cloud~\cite{vatavu2012gestures, vatavu2018q} gestures. Through elicitation, gesture sets have been developed for wheelchair users~\cite{carrington2014gest}, wrist-worn devices~\cite{xu2022enabling}, and augmented reality~\cite{piumsomboon2013user}. In particular, Zhao et al.~\cite{zhao2022above} and Tian et al.~\cite{tian2024sma} elicited above-the-neck gestures and gestures for spinal muscular atrophy respectively. Our work draws inspiration from prior research on the preferences of people with upper-body motor impairments in personalizing mid-air gestures~\cite{yamagami2023people}---even within the relatively specific population of users with motor disabilities, there exists considerable variation in which body parts individuals use to perform upper-body gestures. This diversity highlights the need for personalized input systems that can adapt to and leverage the unique motor capabilities of each user, rather than relying on a one-size-fits-all approach.

\subsection{Virtual Reality Text Entry Methods}
Text entry in VR has been an active research area. A survey by Dube and Arif~\cite{dube2019review} classified more than 30 techniques across 11 categories and concluded that different approaches are effective under different situations: a user might, for example, use a pointing-based technique for short inputs and switch to a physical keyboard for extended writing. We therefore organize prior work by input modality. 

\textit{Physical keyboards} when used in VR achieved the highest reported speeds. Grubert et al.~\cite{grubert2018standard} reported 41.2--43.7 WPM, but required external hardware. \textit{Mid-air hand typing} on virtual keyboards was the most widely studied alternative, including two-finger~\cite{dudley2019envelopes} and ten-finger variants~\cite{yi2015atk, foy2021coactivations}, and pinch selection~\cite{fashimpaur2020pinchtype, bowman2001pinch}. \textit{Word-gesture typing} methods adapted swipe keyboards to VR using 6-DOF controllers~\cite{chen2019word} and force-based digital thimbles~\cite{dube2022shapeshifter}. 
\textit{Gaze- and head-based typing} provided accessible alternatives to hand-based interactions: dwell-based gaze typing~\cite{rajanna2018gaze, majaranta2002twenty} were accelerated through multi-threshold dwell schemes~\cite{mutasim2025dwell} and fused with gesture typing via implicit gaze~\cite{zhao2023gazespeedup}. 
Head-based tapping, dwelling, and gesturing on virtual keyboards~\cite{yu2017tap, xu2019ringtext, lu2020hands} provided additional hands-free options.
\textit{Speech input} were explored both standalone~\cite{pick2016swifter, ruan2018speech} and in combination with a mid-air keyboard to reach 28 WPM~\cite{adhikary2021speech}.

Pointing and other hand-based text entry common in consumer VR has been a barrier for users with upper-limb motor impairments~\cite{pococke2025freehand, mott2020assuming}. Alternative modalities each have their own accessibility trade-offs: gaze typing can be fatiguing and sensitive to calibration~\cite{majaranta2002twenty}, speech input struggles with ambient noise~\cite{derby2020we, sinlapanuntakul2022effects} and impaired speech~\cite{findlater2010beyond, young2010difficulties}, and word-gesture or mid-air techniques rely on sustained arm movements and fine motor control. The modality that is most usable for one person may therefore be unusable for another~\cite{findlater2010beyond}. Our \textit{AdaptType} system instead lets users configure their modality, selection trigger, and entry granularity, making the input choice more accessible and personalized.

%% file: 03-co-design.tex
\section{Co-Design of the \textit{AdaptType} System}
\label{sec:co-design}

Prior work suggested that text entry systems should be customizable~\cite{li2022freedom, mott2020assuming, peracha2026personalized} and multimodal~\cite{li2022freedom, findlater2010beyond, suhm2001multimodal}, yet \textit{how} such customization may be achieved to accommodate a range of motor abilities remains unclear. Prior work also elicited gesture preferences of people with upper-body motor impairments~\cite{yamagami2023people}, but did not establish which gestures might be feasible for extended VR text entry sessions. We therefore developed \textit{AdaptType} through a co-design process with
three people with upper-body motor impairments. 

\subsection{Co-Design Process}

We began by building an initial prototype of \textit{AdaptType} based on prior work insights~\cite{li2022freedom, findlater2010beyond, mott2020assuming, yamagami2023people}. The initial prototype integrated gaze, hand, and voice input, and implemented two gestures for users with limited fine/gross motor function: a \textit{finger pinch} gesture that required no gross motor function, and a \textit{hand wave} gesture that required no fine motor function, based on personalized gestures elicited in prior work~\cite{yamagami2023people}. This prototype allowed us to probe the open design questions through hands-on sessions. 

We then worked with three participants with upper-body motor impairments to co-design the \textit{AdaptType} system. The three participants were all women, on average 70.3 years old ($SD=8.0$), and had varying fine and gross motor functions.\footnote{One participant reported having moderate to severe impairments with gross and fine motor function on one side, and mild impairments on the other; one participant reported having severe bilateral fine motor impairments, and unimpaired gross motor function; one participant reported having mild to moderate gross and fine motor function on both sides.} In a hands-on session, each participant configured a text entry method by working through the customization options, then entered at least 10 phrases from the MacKenzie and Soukoreff phrase set~\cite{mackenzie2003phrase} using their customized technique. Participants also explored state-of-the-art VR text entry techniques such as using a hand raycast and a pinch trigger, when feasible given their motor abilities. Participants then shared their experience and feedback, and brainstormed what their ``ideal keyboard'' would be based on the experience. 

\subsection{Design Considerations}

Our co-design work helped reveal three primary design considerations, described below.

\paragraph{\textbf{Custom gestures should be performable with the hand/finger supported}}
\label{sec:preliminary-finding-1}
Our initial prototype used a \textit{hand wave} gesture in which participants move the whole forearm to trigger selection~\cite{yamagami2023people}. Participants confirmed the hand wave required no fine motor control, but reported fatigue within a few trials. One participant mentioned ``\textit{it was much easier to control the cursor with my hand resting on the mouse and sliding on the desktop}.'' One participant used one hand to hold the other wrist in mid-air while exploring the hand-based technique, and reported it made the gesture easier by ``\textit{stabilizing [their] wrist}.'' The findings suggest that customized gestures for VR text entry should not only consider whether gross/fine motor function is \textit{required}, but also whether the user's hand/finger can \textit{rest} in a comfortable position while performing it. Therefore, we redesigned the custom triggers around movements performed against a supporting surface---a desktop, armrest, or the user's lap (Section~\ref{sec:selection-trigger}). 

\paragraph{\textbf{Custom gestures should support a flexible range of motion}}

Although \textit{finger pinch} was technically feasible for users with mild fine motor impairments, one participant still found this gesture difficult to use, as gestural detection failed when their thumb and index finger were not ``\textit{wide open}.'' They also reported fatigue with their fingers after usage. This finding indicated that custom gestures should support a flexible range of motion, and informed our design of calibrating custom pointers and triggers to each user's range of motion. In the iterated design, a user can lightly lift up their palm or finger, curl up their wrist, or use other gestures that work best with their motor function. 

\paragraph{\textbf{Effective customization requires scaffolding beyond enumeration}}
\label{sec:preliminary-finding-2}
During the sessions, all participants spent over an hour exploring all customization options and configuring their personalized input method. Participants found the process ``\textit{overwhelming},'' and repeatedly asked to revisit settings as they could no longer clearly recall which ones had worked well for them. This finding indicated that enumeration would fail when the customizable system offers too many available options, and motivated our design of narrowing the choice and offering ability-based recommendations as a starting point (Section~\ref{sec:ability-based-recommendations-of-customization-settings}), rather than presenting users with the full set of options upfront.

%% file: 04-system-design.tex
\section{AdaptType}
\label{sec:system-design}

In this section, we present the design and implementation of the \textit{AdaptType} system in two main respects: (1) the end-user customization framework, and (2) the scaffolded customization process based on users' motor abilities. 

\begin{figure*}[h]
    \vspace{-1pc}
    \includegraphics[width=0.88\textwidth]{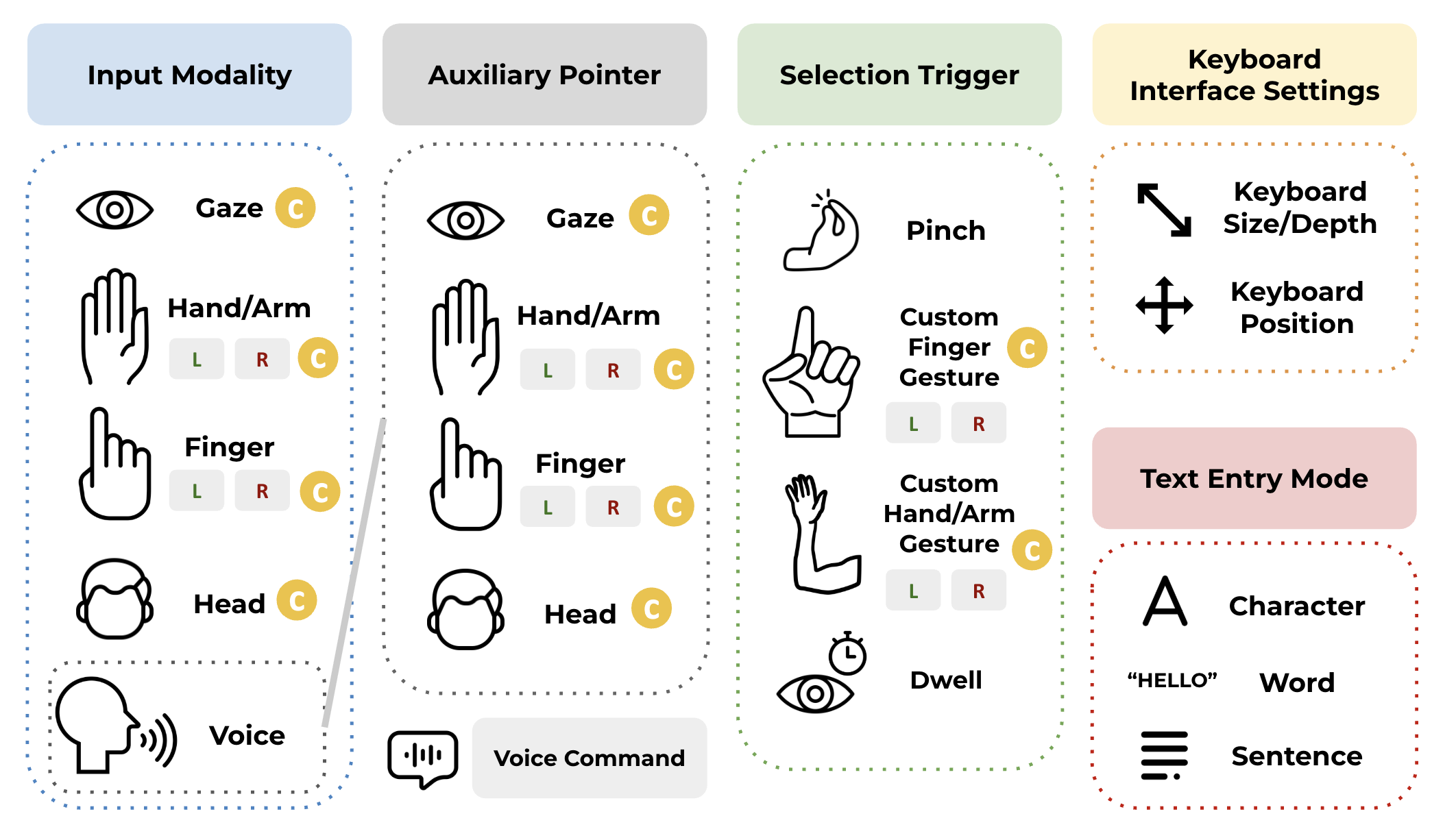}
    \vspace{-1pc}
    \caption{The AdaptType end-user customization framework. There are five categories of customization settings: (1) input modality, (2) auxiliary pointer (when the input modality is voice), (3) selection trigger, (4) keyboard interface settings, and (5) text entry mode. Voice command is also included as a stand-alone feature. The ``L'' and ``R'' labels represent the handedness of modalities, auxiliary pointers, and selection triggers. The ``C'' labels indicate that the modalities, auxiliary pointers, and selection triggers can be calibrated to match the user's abilities, e.g., posture or range of motion.}
    \vspace{-0.5pc}
    \Description{Overview of the AdaptType end-user customization framework. Customization is organized into five categories: input modality, with gaze, hand/arm, finger, head, and voice options; auxiliary pointer, with gaze, hand/arm, finger, and head options; selection trigger, with pinch, custom finger gesture, custom hand/arm gesture, and dwell options; keyboard size/depth and position; and character-, word-, or sentence-level text entry modes. Voice commands provide an additional control mechanism. Left/right indicators denote handedness, and calibration indicators identify techniques that can be calibrated to users' abilities.}
    \label{fig:end-user-customization}
\end{figure*}

\subsection{The End-User Customization Framework}

We include a summary of all the customization options \textit{AdaptType} supports in Figure~\ref{fig:end-user-customization}. There are five categories of customization settings: (1) input modality, (2) auxiliary pointer (when the input modality is voice), (3) selection trigger, (4) keyboard interface settings, and (5) text entry mode. We conceptualized the input modality, auxiliary pointer, and selection trigger following anatomies of 3D pointing in prior work~\cite{pfeuffer2017gaze+, wagner2023fitts}.

\subsubsection{Setting 1: Input Modality}
\label{sec:input-modality}

The ``input modality'' refers to the way a user indicates their desired text to input, whether by pointing at a visual target, or by speaking using voice recognition.

\paragraph{\textbf{Gaze}} 
Users look at the keys to enter text. The gaze input modality is implemented using the Meta OVREye library to obtain the user's gaze direction in 3D world coordinates.

\paragraph{\textbf{Hand/Arm (Gross Motor Function)}} 
Users control a hand-based pointer emitted from the palm using their left/right hand or forearm to point at the keys. This modality is implemented using the Meta OVRHand library to obtain real-time position and direction of the palm, from which a hand-based raycast is computed. Users may manipulate the pointer using wrist or forearm movements, requiring only gross motor control. By default, the hand raycast directly maps the palm's six degrees-of-freedom (DoF) position and direction to the pointer's origin and direction. However, certain regions of the keyboard may be difficult or unreachable for some users under this mapping. To address this, \textit{AdaptType} provides a calibration mechanism that allows users to draw a rectangular area in mid-air that they are comfortable reaching with their hand and forearm. The system then computes a mapping from the detected palm pose to the raycast such that the full keyboard remains reachable while the user's hand stays within their specified area.

\paragraph{\textbf{Finger (Fine Motor Function)}} 
Users control a finger-based pointer emitted from their left/right index finger to point at the keys. This modality is implemented using the Meta OVRHand and OVRSkeleton libraries to obtain real-time hand joint positions, from which a finger-based raycast is computed. Users manipulate the pointer through index finger movements, relying on fine motor control without requiring gross arm movement. \textit{AdaptType} provides a similar calibration mechanism as for hand/arm pointing that enables access to the full keyboard. 

\paragraph{\textbf{Head}} 
Users control a head-based pointer emitted between the eyes and in the head's forward orientation to point at the keys. This modality is implemented using the Meta OVREye library to obtain real-time eye anchor positions, from which a head-directed raycast is computed. The head pointer is calibrated using similar approaches as the hand and finger pointers to address challenges with reachability. 

\paragraph{\textbf{Voice}} 
Users dictate the text they wish to enter. This modality is implemented using the NativeWebSocket open source library\footnote{https://github.com/endel/NativeWebSocket} and OpenAI's real-time transcription API,\footnote{https://developers.openai.com/api/docs/guides/realtime-transcription} using the \texttt{gpt-4o-mini-\\transcribe} model. Transcribed speech is treated as input text once a complete transcription segment is returned by the API.

\subsubsection{Setting 2: Auxiliary Pointer}

The ``auxiliary pointer'' refers to the way a user indicates a visual target when their primary input modality is voice, when there is no pointer that allows for target pointing and selections. Thus, we offer an \textit{auxiliary pointer} option including a similar set of options as in Section~\ref{sec:input-modality}: gaze, hand/arm, finger, and head. 

\subsubsection{Setting 3: Selection Trigger}
\label{sec:selection-trigger}

The ``selection trigger'' refers to the way a user sends a selection event to the system to actuate whatever they are indicating with their ``input modality.''

\begin{figure}[h]
    \includegraphics[width=\linewidth]{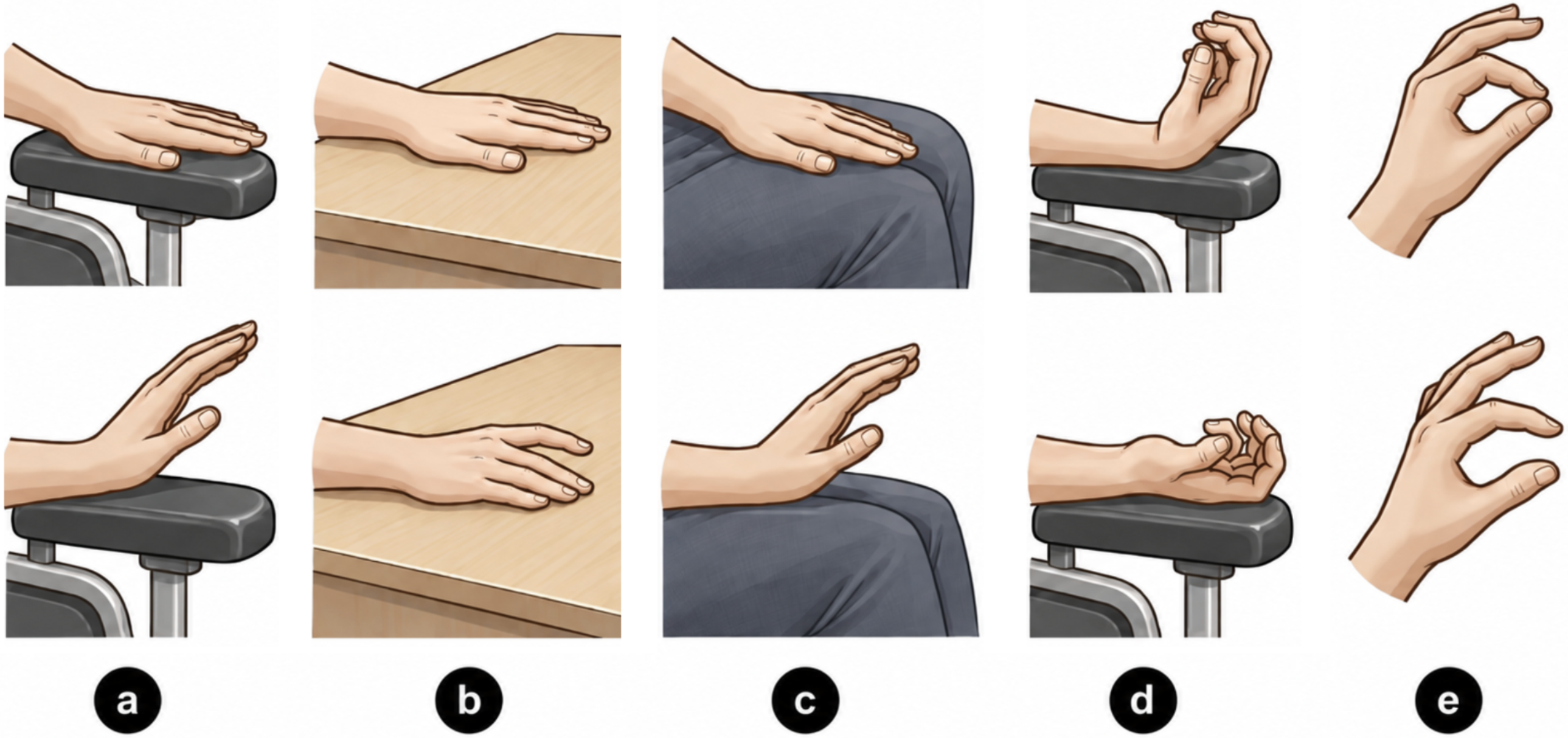}
    \vspace{-1.5pc}
    \caption{A non-exhaustive set of example trigger gestures a user can select. (a) Hand tap on armrest; (b) Finger tap on desktop; (c) Hand tap on lap; (d) Wrist curl; (e) Pinch.}
    \vspace{-1pc}
    \Description{Five illustrated examples of selection-trigger gestures supported by AdaptType, shown in resting and activated positions: a hand tap on a wheelchair armrest, a finger tap on a desktop, a hand tap on the user's lap, a wrist curl performed on an armrest, and a finger pinch. The examples illustrate how gestures can be performed with the hand or arm supported on different surfaces.}
    \label{fig:trigger-gestures}
\end{figure}

\paragraph{\textbf{Pinch}} Users close their thumb and index finger to perform a ``pinch'' gesture, and open their fingers to deactivate the trigger (Figure~\ref{fig:trigger-gestures}d). The pinch gesture is detected using the Meta OVRHand library. This selection trigger requires high finger dexterity; meanwhile, this trigger can be performed on the desktop, armrest, user's lap, or any position visible to the headset, thus requiring little to no gross motor control.

\paragraph{\textbf{Custom Finger Gesture}} 
Users perform a custom finger gesture using their index finger; for example, by tapping on a surface such as a desktop, armrest, or lap (Figure~\ref{fig:trigger-gestures}c), while the hand and wrist remain supported. This method derives an implicit index-finger raycast from hand joint positions and triggers selection when the raycast's vertical angle exceeds a threshold, with a 100~ms smoothing window applied. The default threshold is set to zero, assuming a flat surface. To calibrate this gesture, users perform the gesture (e.g., tapping) $10$ times, and the system records the range of vertical motion of the implicit finger raycast, setting the trigger threshold to the median of this range. 

\paragraph{\textbf{Custom Hand/Arm Gesture}} Users perform a custom hand/arm gesture by moving their entire hand or forearm; for example, by tapping on a surface such as a desktop, armrest, or lap (Figure~\ref{fig:trigger-gestures}a). The custom hand/arm gesture is calibrated using a similar approach as the custom finger gesture. 

Note that although the implementation of custom finger and hand/arm gestures are based on heuristics, the detection mechanism supports a surprisingly rich set of custom gestures without requiring resource-intensive model training. For example, aside from the ``finger tap'' (Figure~\ref{fig:trigger-gestures}c) and ``hand tap'' (Figure~\ref{fig:trigger-gestures}a) gestures as default options, users can perform a wrist curl (Figure~\ref{fig:trigger-gestures}b), a finger curl, a hand flip, and many other gestures that align with their motor abilities, as long as the vertical movement is detectable. 

\paragraph{\textbf{Dwell}} 
Users trigger selection by dwelling on a target. The default dwell time is 600~ms, a duration that balances speed, accuracy, and usability as prior work recommends~\cite{paulus2021usability}.

\paragraph{\textbf{Voice Command}} We also include a ``voice command'' feature as a stand-alone control mechanism. Unlike pointer-based inputs, voice commands do not function as selection triggers but instead directly manipulate text and interface elements. When voice command is enabled, the system sends the transcribed text with the current input box content to a GPT-4o-mini endpoint to determine whether the input corresponds to a command (e.g., ``enter,'' ``clear,'' or ``change the first word to chocolate'') or text entry. The system then executes the inferred command or updates the text accordingly.

\subsubsection{Setting 4: Keyboard Size and Position}

Users can adjust the keyboard size/depth and position through voice control such as ``move keyboard up,'' ``move keyboard left,'' ``move keyboard closer,'' ``move keyboard farther,'' etc. The keyboard depth control is restricted within a bound so that the keyboard does not exceed $\pm35^\circ$ horizontally and vertically from the center of view, a comfortable range within a user's view \cite{burlingham2024motor}, and that each key has minimum width of $5^\circ$ to ensure gaze pointing usability. 

\subsubsection{Setting 5: Text Entry Mode} 
Using the selected input modality and selection trigger, users can enter text character by character by pointing at the keys. Alternatively, a swipe-based method is available in which users maintain the trigger in an active state while moving across keys and release the trigger to input a word.\footnote{During the co-design sessions (Section~\ref{sec:co-design}), this method was found to be less usable, with higher cognitive load and a steeper learning curve. As a result, it is included as an exploratory option but is not among the recommended configurations in Section~\ref{sec:ability-based-recommendations-of-customization-settings}.} When voice is selected as the input modality, text is entered at word/phrase-level, while character-level editing remains available in parallel.

\subsubsection{Text Entry Interface Visualization and Feedback Implementation} 
\begin{figure}[h]
    \vspace{-0.5pc}
    \includegraphics[width=\linewidth]{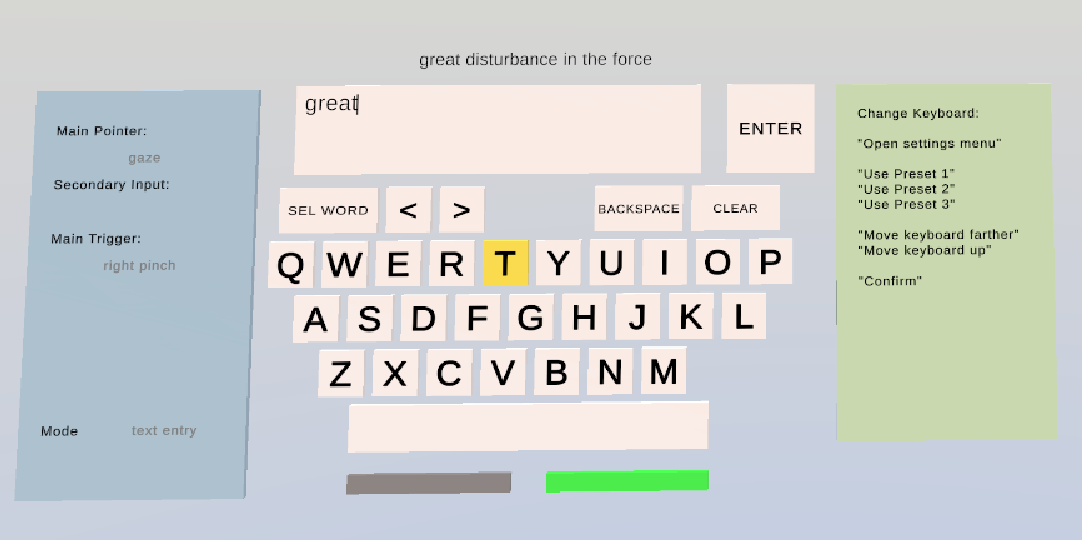}
    \vspace{-1pc}
    \caption{The \textit{AdaptType} text entry interface. The keyboard builds on top of the standard QWERTY keyboard layout with function keys at the top of the keyboard: ``sel word'' to select the word at the current cursor location, ``<'' and ``>'' to move the cursor to the left/right, ``backspace'' to delete a character or the selected text, and ``clear'' to clear the input text box. Users can also place the cursor using their selected pointer to point at the intended location, then activating the selection trigger (for example, looking at the letter ``t'' and performing a pinch). The left information panel shows the currently selected customization options, and the right panel displays hints for keyboard voice commands. The user is currently looking at the letter T and performing a pinch gesture with their right hand, as indicated by the right green status bar.}
    \vspace{-1pc}
    \Description{Screenshot of the AdaptType VR text entry interface. A QWERTY keyboard appears between a left panel showing the current customization settings and a right panel showing voice-command hints. Function keys above the keyboard support word selection, cursor movement, backspace, and clearing text. In the example, the user is looking at the letter T and activating a right-hand pinch trigger, indicated by a highlighted key and green trigger-status bar.}
    \label{fig:keyboard-layout}
\end{figure}

\begin{figure*}[]
    \vspace{-0.5pc}
    \includegraphics[width=\textwidth]{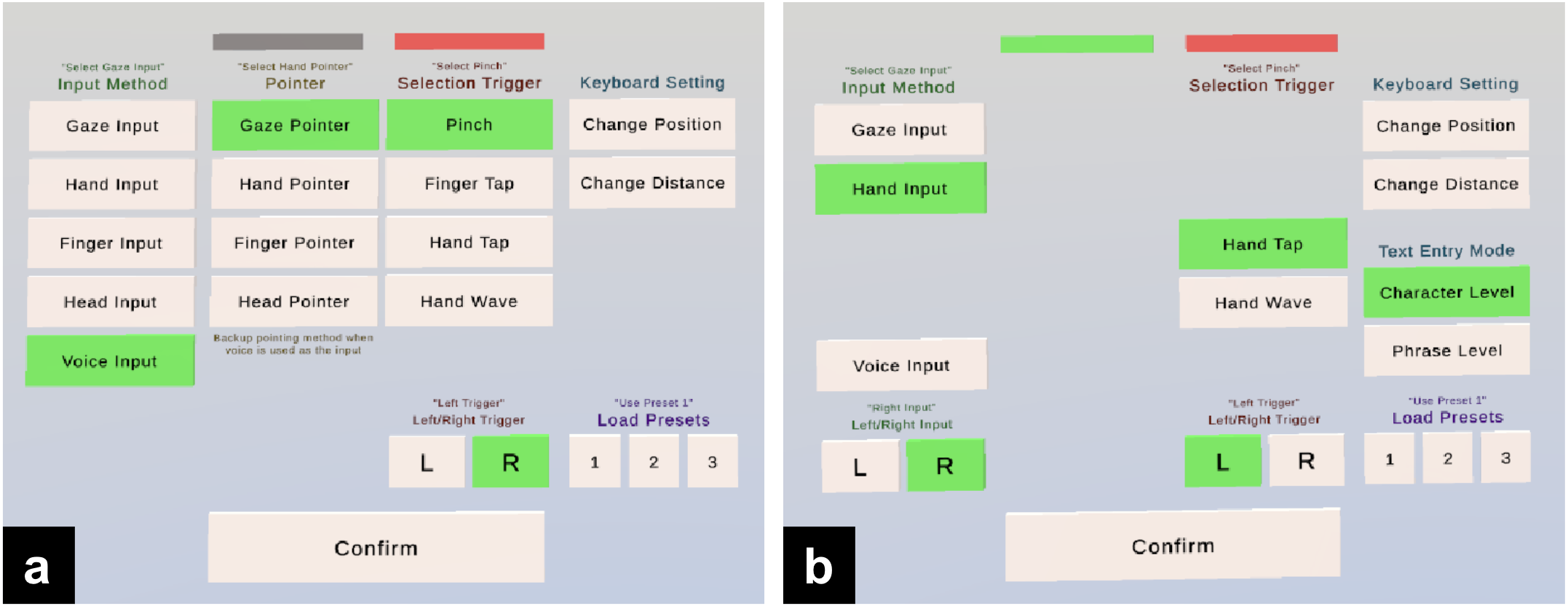}
    \vspace{-1.5pc}
    \caption{The settings menu of AdaptType, which allows users to customize their text entry method: (a) an example settings menu displaying most of the available customization options for a user with less severe motor impairments; (b) an example settings menu displaying the available customization, for example, when a user does not have fine motor function and has impaired neck movement.}
    \vspace{-1pc}
    \Description{Two screenshots of the AdaptType customization menu illustrating ability-based filtering of available options. Panel a shows a relatively broad set of input modalities, auxiliary pointers, selection triggers, keyboard settings, and text entry modes for a user with less severe motor impairments. Panel b shows a reduced set of options for a user without fine motor function and with impaired neck movement, illustrating how inaccessible options are removed while suitable alternatives remain.}
    \label{fig:config-examples}
\end{figure*}

The text entry interface is positioned at the center of a vertical 2D task plane five meters away from the user, with a keyboard in the middle and left and right panels showing configuration information and usage tips, respectively (Figure~\ref{fig:keyboard-layout}). The hand- and finger-based pointers are visualized as red beams emitted from the user’s hand or finger toward the keyboard, and keys are highlighted in yellow when hovered over, similar to state-of-the-art VR pointing visualizations. The raycast beam is not visualized for gaze or head input, as visible cursors can degrade gaze performance due to cursor chasing~\cite{majaranta2002twenty}. Selection triggers are visualized using status bars, which appear gray when inactive and green when activated (Figures~\ref{fig:keyboard-layout} and ~\ref{fig:config-examples}a), except dwell, where the selected key turns green from the yellow hover highlight to indicate activation.

\subsection{Ability-Based Recommendations of Customization Settings}
\label{sec:ability-based-recommendations-of-customization-settings}

While \textit{AdaptType} offers a rich set of customization options that enable users to select input methods aligned with their abilities (Figure~\ref{fig:config-examples}a), the breadth of these options may be overwhelming. Users may also be unaware of the full range of options or which options best suit their needs~\cite{findlater2007evaluating, findlater2008impact, kong2025supporting}. To address this challenge, \textit{AdaptType} guides customization through an ability-based recommendation of settings leveraging large-landuage models (LLMs). 

\paragraph{\textbf{Requesting Recommendations from LLMs}} 
We abstract users' anonymized responses describing specific aspects of their motor abilities from the screening questionnaire, including condition descriptions, $11$ ratings using the QuickDASH questionnaire~\cite{gummesson2006shortened}, four ratings of lateralized motor abilities (left/right, fine/gross), one rating of neck movement, and one rating of speaking ability. A request is then sent to a \texttt{gpt-4o-mini} endpoint to generate five recommended combinations in the format \texttt{<InputModality|\\SelectionTrigger>}, with voice excluded from the recommendations. Upon receiving the response, the system parses the recommended combinations and iterates through them to identify the first combination in which the input modality and selection trigger are compatible (e.g., a user cannot use the same hand to manipulate the pointer and perform a tapping selection gesture).

\paragraph{\textbf{Propagating Recommended Presets}} 

The \textit{AdaptType} system offers three presets for each user to explore. Note that the presets do not necessarily represent what is likely the \textit{best} option for the user; instead, they aim to span a variety of customization options that may be feasible for the user, without overwhelming the user with the number of options.
The first valid \texttt{<InputModality|\\SelectionTrigger>} pair parsed from the GPT response is used as \textbf{Preset~1}, emphasizing manual control. The system then iterates through the selection triggers compatible with the user's abilities to identify an alternative trigger (\texttt{AltTrigger}) that differs from the initial recommendation. This trigger is paired with the gaze input modality as the LLM recommendation prioritizes hand- or finger-based modalities unless motor impairments are severe. Thus, gaze input and \texttt{AltTrigger} form \textbf{Preset~2}, which emphasizes alternative input and triggering strategies. For Preset~3, the system selects voice as the input modality and identifies a third compatible selection trigger,\footnote{The same selection trigger will be used if there is no more available option.} along with an \texttt{AltPointer} not used in Presets 1 or 2 as an auxiliary input. Voice input, the \texttt{AltPointer}, and the selected trigger together form \textbf{Preset~3}. Collectively, these three presets span a range of interaction strategies aligned with the user's motor abilities, from manual pointer-based control to voice-driven input.

\paragraph{\textbf{Identifying Available Customization Options}} 
In addition to the LLM-based recommendations, \textit{AdaptType} applies a heuristic-based filter to exclude customization options that are likely inaccessible given the user's motor abilities. Input modalities and selection triggers involving the hand, finger, or head are removed if the user indicates an ``unable'' ($5/5$) rating for the corresponding motor ability. Furthermore, head- and finger-based input modalities and auxiliary pointers are excluded when users report more than moderate impairment ($>3/5$), as these interactions require finer motor control or impose greater strain on the neck when using a head-mounted display.

\paragraph{\textbf{End-User Customization through Voice Commands}} 
Based on the recommendations, users can further configure their text entry method using conversational interactions. Users can say ``open settings menu'' to access the configuration interface (Figure~\ref{fig:config-examples}) and ``confirm'' to apply the current selection and return to the text entry interface. Within the settings menu, users can issue commands such as ``select voice input'' or ``select pinch trigger'' to modify settings. 

%% file: 05-method.tex
\section{Study Method}

We conducted a lab study with 10 participants to evaluate \textit{AdaptType}. The goal of the study was to validate the effectiveness of \textit{AdaptType} in enabling text entry configurations that matched user abilities, to understand users' customization behaviors and preferences, and to investigate whether customized multimodal techniques improved the accessibility of VR text entry. 

\begin{table*}[t]
\vspace{-0.5pc}
\centering
\begin{minipage}[t]{0.48\textwidth}
\vspace{0pt}
\centering
\resizebox{\linewidth}{!}{%
\begin{tabular}{@{}cccl@{}}
\toprule
\textit{\textbf{ID}} &
  \textit{\textbf{Gender}} &
  \textit{\textbf{Age}} &
  \multicolumn{1}{c}{\textit{\textbf{Condition}}} \\ \midrule
\textbf{P1}  & Man   & 51 & Spinal cord injury                                      \\ \midrule
\textbf{P2}  & Man   & 69 & Shoulder pain                                           \\ \midrule
\textbf{P3}  & Man   & 23 & Spinal cord injury                                      \\ \midrule
\textbf{P4}  & Woman & 71 & \begin{tabular}[c]{@{}l@{}}Autonomic small fiber neuropathy,\\ cervical radiculopathy\end{tabular} \\ \midrule
\textbf{P5}  & Woman & 60 & Amputation                                              \\ \midrule
\textbf{P6}  & Woman & 67 & Facioscapulohumeral muscular dystrophy                  \\ \midrule
\textbf{P7}  & Man   & 38 & Facioscapulohumeral muscular dystrophy                  \\ \midrule
\textbf{P8}  & Man   & 63 & Amputation                                              \\ \midrule
\textbf{P9}  & Woman & 59 & Spinal cord injury                                      \\ \midrule
\textbf{P10} & Woman & 63 & Spinal cord injury                                      \\ \bottomrule
\end{tabular}%
}
\vspace{0.5pc}
\caption{Participant demographics. \textit{Condition} indicates each participant's self-reported physical condition influencing their upper-body motor ability.}
\Description{Demographic characteristics of the 10 study participants, listing participant ID, gender, age, and self-reported condition affecting upper-body motor ability. Participants include five women and five men, ages 23 to 71, with conditions including spinal cord injury, amputation, facioscapulohumeral muscular dystrophy, shoulder pain, autonomic small fiber neuropathy, and cervical radiculopathy.}
\label{tab:participant-demographics}
\end{minipage}%
\hfill%
\begin{minipage}[t]{0.47\textwidth}
\vspace{0pt}
\centering
\includegraphics[width=\linewidth]{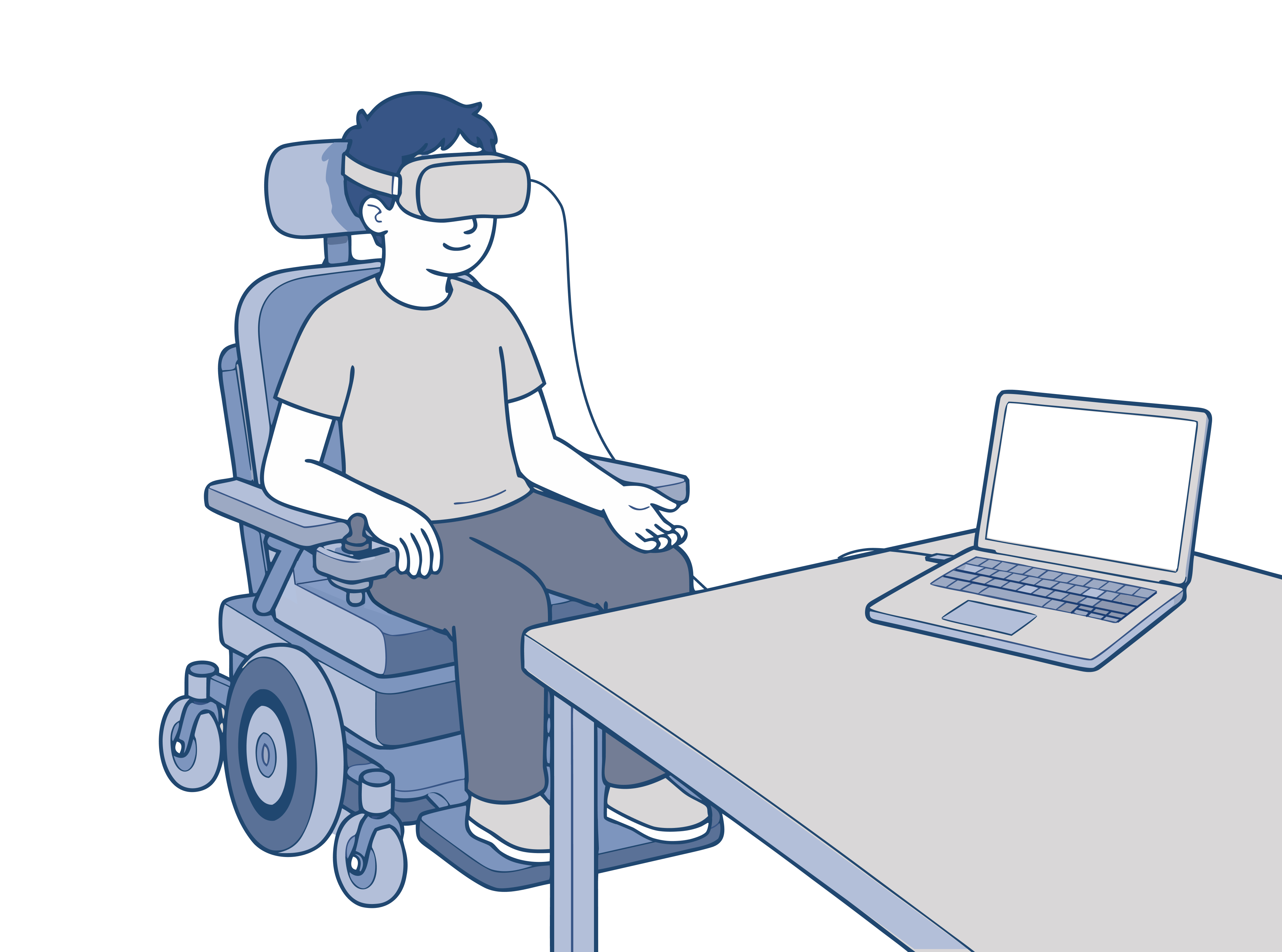}
\vspace{0pc}
\makeatletter\def\@captype{figure}\makeatother
\caption{Illustration of the study setup. A participant seated at a desk wearing the Meta Quest Pro headset, connected to an experiment laptop facing the experimenter.}
\Description{Illustration of the study apparatus and physical setup. A participant is seated at a desk wearing a Meta Quest Pro headset, with the headset connected to the experiment laptop on the desk. The participant's hands rest in a supported position while using the VR system.}
\label{fig:experiment-setup}
\end{minipage}
\vspace{-1.5pc}
\end{table*}

\subsection{Participants}

We recruited 10 participants with upper-body motor impairments (Table~\ref{tab:participant-demographics}) through email solicitations, organizational newsletters, and snowball sampling. On average, participants were 56.4 years old ($SD=15.2$), with 5 women and 5 men. Reported health conditions included spinal cord injury (4), amputation (2), facioscapulohumeral muscular dystrophy (2), shoulder pain (1), and autonomic small fiber neuropathy and cervical radiculopathy (1). Participants reported having moderate expertise with text entry using physical keyboards ($M=3.0, SD=1.3$, on a novice--expert scale of 1--5), virtual keyboards ($M=3.1, SD=1.5$), and voice input ($M=3.3, SD=1.5$), and having novice-level experience with VR ($M=1.2, SD=0.6$). Participants were compensated \$40 per hour, pro-rated to each half-hour, plus transportation costs. 

\subsection{Apparatus}

We used a Meta Quest Pro headset ($1800 \times 1920$ px per eye, 90~Hz refresh rate, $106^{\circ} \times 96^{\circ}$ FoV) for our study, with 72~Hz eye tracking rate and 60~Hz hand tracking rate. The \textit{AdaptType} prototype was a Unity app (Figure~\ref{fig:keyboard-layout} and~\ref{fig:config-examples}, version 6000.1.2f1) built on a Windows 11 laptop with a 2.4~GHz Intel i7 CPU and GeForce GTX 4060 GPU, running on the Meta Quest Pro headset via a Meta Link cable (Figure~\ref{fig:experiment-setup}). Study sessions were conducted in a lab setting, with participants seated at a desk (Figure~\ref{fig:experiment-setup}). 

\subsection{Study Procedure}
\label{sec:procedure}

Study sessions were conducted in person with approval from our university’s Institutional Review Board (IRB). Participants completed a screening survey including demographic questions and 17 questions on their motor and speaking abilities (11 questions of the QuickDASH questionnaire~\cite{gummesson2006shortened}, with an additional four questions on lateral fine/gross motor abilities, one question on neck movement, and one question on speaking ability), whose responses were used to generate ability-based recommendations. After a brief introduction, participants were fitted with the headset and completed an eye-tracking calibration procedure through the Meta Quest Pro calibration dialog. Each session lasted approximately two hours, and short breaks were offered throughout the session. 

\paragraph{\textbf{Part 1: Initial Customization}} Participants were guided to use \textit{AdaptType} to select their customized text entry technique, based on the three ability-based recommended keyboard presets (Section~\ref{sec:ability-based-recommendations-of-customization-settings}). Participants completed a calibration process for each new custom trigger/pointer, and transcribed one sentence with each recommendation. Participants were then instructed to open the settings menu (Figure~\ref{fig:config-examples}) and freely explore other available options.\footnote{Although \textit{AdaptType} fully supports conversational end-user customization, an experimenter-controlled interface was used as a complementary method to switch between pointers and triggers. This served as a Wizard-of-Oz mechanism to expedite later-stage customization, thereby avoiding excessively long study sessions.} Participants could adjust their customization choices until they were satisfied with their configuration after successfully completing one full sentence with the selected technique.

\newlength{\namew}\newlength{\descw}
\begin{table*}[h]
\vspace{-0.5pc}
\centering
\small
\setlength{\tabcolsep}{4pt}
\renewcommand{\arraystretch}{1.05}
\begin{minipage}[t]{0.37\textwidth}
\vspace{0pt}
\settowidth{\namew}{\textbf{Technique}}%
\setlength{\descw}{\linewidth}\addtolength{\descw}{-\namew}\addtolength{\descw}{-2\tabcolsep}%
\begin{tabular}[t]{@{}l p{\descw}@{}}
\toprule
\multicolumn{2}{@{}p{\linewidth}@{}}{\raggedright Text Entry \textbf{Techniques}\newline Participants Used} \\
\midrule
\textbf{Technique} & \raggedright \textbf{Description} \tabularnewline
\midrule
\parbox[t]{\namew}{\raggedright\textit{Custom-}\newline\textit{Initial}} & \raggedright
  The participant's personalized \mbox{\textit{AdaptType}} configuration selected in Part 1. \tabularnewline \addlinespace[5pt]
\textit{Gaze} & \raggedright
  Gaze for pointing with dwell selection. \tabularnewline \addlinespace[5pt]
\textit{Hand} & \raggedright
  A hand raycast for pointing, with a pinch selection trigger.\footnotemark{} \tabularnewline \addlinespace[5pt]
\textit{Voice} & \raggedright
  Dictation for text entry and voice commands for editing, with no keyboard or pointer. \tabularnewline
\bottomrule
\end{tabular}
\end{minipage}%
\hfill%
\begin{minipage}[t]{0.60\textwidth}
\vspace{0pt}
\settowidth{\namew}{\textit{Transcription}}%
\setlength{\descw}{\linewidth}\addtolength{\descw}{-\namew}\addtolength{\descw}{-2\tabcolsep}%
\begin{tabular}[t]{@{}l p{\descw}@{}}
\toprule
\multicolumn{2}{@{}p{\linewidth}@{}}{\raggedright Text Entry \textbf{Tasks}\newline Completed with Each Technique} \\
\midrule
\textbf{Task} & \raggedright \textbf{Description} \tabularnewline
\midrule
\textit{Editing} & \raggedright
  Participants corrected a partially entered phrase to match a target. Errors were seeded by perturbing the target at the character level: for a phrase of length $N$, between 5\% and 15\% of the characters were selected at random and modified by substitution, deletion, or insertion, each with equal probability. Only lowercase letters were substituted or inserted. This yielded 1--4 edits for phrases of approximately 30 characters. Participants completed two practice trials, then a two-minute test block. \tabularnewline \addlinespace[5pt]
\textit{Transcription} & \raggedright
  Participants entered complete target phrases from scratch. Participants completed two practice trials, then a five-minute test block.  
  \tabularnewline
\bottomrule
\end{tabular}
\end{minipage}
\vspace{0.2pc}
\caption{The four text entry techniques compared in Part 2 of the study (left) and the two tasks participants completed with each technique (right). Technique order was counterbalanced using a balanced Latin square~\cite{williams1949experimental}. Target phrases in both tasks were randomly drawn from the MacKenzie and Soukoreff phrase set~\cite{mackenzie2003phrase}. Participants were instructed to keep going until signified to stop, when (1) they had reached the test block time bound (2 minutes for editing, 5 minutes for transcription), and (2) they had completed at least five trials. Participants were also advised to proceed to the next trial if one editing or transcription trial was taking longer than 60 or 90 seconds, respectively. We used this method to ensure reasonable study duration in cases of excessively long trials.}
\Description{Study techniques and tasks used in Part 2. The four text entry techniques are each participant's Custom-Initial AdaptType configuration, Gaze with dwell selection, Hand with hand-ray pointing and pinch selection, and Voice with dictation and voice commands. The two tasks are Editing, in which participants correct artificially seeded character-level errors in an existing phrase during a two-minute block, and Transcription, in which participants enter complete target phrases during a five-minute block. Technique order was counterbalanced, and both tasks used phrases from the MacKenzie and Soukoreff phrase set.}
\label{tab:techniques-and-tasks}
\vspace{-1.5pc}
\end{table*}
\footnotetext{The \textit{Hand} technique was omitted for the six participants without the fine and gross motor control it requires.}

\paragraph{\textbf{Part 2: Text Entry Tasks}} Participants then completed text entry tasks using four techniques (Table~\ref{tab:techniques-and-tasks}, left) $\times$ two tasks (Table~\ref{tab:techniques-and-tasks}, right). After completing the tasks for each technique, participants completed the NASA-TLX questionnaire~\cite{hart1988development, hart2006nasa} and a usability questionnaire adapted from the System Usability Scale~\cite{brooke1996sus}, respectively for the editing and the transcription task.   

\paragraph{\textbf{Part 3: Iterated Customization}} If a participant completed Parts 1 \& 2 in less than two hours and reported no significant fatigue, they also then completed a round of iterated customization. Participants were given the option to change their customization settings and adopt features that they found useful based on their experiences in Parts 1 \& 2, yielding the \textit{Custom-Iterated} technique. Participants then used this technique to complete the transcription task and the NASA-TLX and usability questionnaires. Participants also shared additional thoughts on their experience.\footnote{During the first two sessions, we observed high voice recognition error rates among older adult participants using the initial Meta Voice SDK implementation. We therefore updated the voice recognition pipeline to use OpenAI's real-time transcription API with the \texttt{gpt-4o-mini-transcribe} model. This change was implemented after P2, and the first two participants did not complete the iterated customization and associated text entry tasks.}

\subsection{Data}

We collected a total of 782 trials from the 10 participants, including 244 trials using the \textit{Custom-Initial} text entry technique, 119 trials using the \textit{Gaze} technique, 51 trials using the \textit{Hand} technique, 257 trials using the \textit{Voice} technique, and 111 trials using the \textit{Custom-Iterated} technique. There were 184 total trials collected from the editing task and 598 total trials collected from the transcription task. We excluded P1 and P2's data for the \textit{Voice} technique from the analyses, as the voice input implementation was updated after P2, thereby affecting the quality of voice command processing. The \textit{Hand} technique was omitted for P1, P3, P6, P8, P9 and P10, who did not have the fine and gross motor control required to use it. Among the collected data, 17 trials were failed attempts (2.2\%),\footnote{A trial was marked as a failed attempt if the participant attempted to enter the target phrase, but (1) aborted without submitting, as they were too frustrated or the task was taking too long, or (2) ended up clearing the text box and submitting with no text.} which we excluded from the text entry performance analyses of speed, accuracy, and throughput, leaving 765 trials. The \textit{Custom-Iterated} technique was collected only for P3--P8, and only for the transcription task.

\subsection{Measures} We measure the text entry experience using metrics in three respects: (1) performance, (2) perceived task load, and (3) usability ratings. For transcription tasks, we report standard text entry metrics: speed (WPM)~\cite{mackenzie2002note}, corrected error rate (CER), uncorrected error rate (UER), and total error rate (TER)~\cite{soukoreff2003metrics,zhang2019beyond}. We additionally report \textit{throughput}, which combines speed and accuracy into a single measure of efficiency~\cite{zhang2019text}. For editing tasks, we report \textit{duration} (the elapsed time from the first input event to trial confirmation) and the UER of the final edited sentence relative to the target phrase. Participants filled out the NASA Task Load Index~\cite{hart1988development, hart2006nasa} for each technique, which captures mental demand, physical demand, temporal demand, performance, effort, and frustration. We additionally administered a five-item usability questionnaire covering ease of use, ease of learning, comfort, fatigue, and anticipated future use. All items used a 1--10 scale, where 1 indicated the most positive rating and 10 the least. 

\subsection{Analysis} 

For transcription speed (WPM) and editing duration, we ran linear mixed models (LMM) \cite{west2015linear, littell1998statistical} with \textit{Technique} as a fixed factor and \textit{Subject} and \textit{Phrase} as random factors. For throughput, which had a single value per participant $\times$ technique, we ran an LMM with \textit{Technique} as the fixed factor and \textit{Subject} as a random factor.
For accuracy, we modeled UER, CER, and TER as error counts over a total character denominator \cite{soukoreff2003metrics}, fitting beta-binomial generalized linear mixed models (GLMM) \cite{brooks2017glmmtmb} with \textit{Technique} as a fixed factor and \textit{Participant} and \textit{Phrase} as random factors. We ran mixed ordinal logistic regression models \cite{hedeker1994random} to compare task load and usability ratings, modeling \textit{Technique} as a fixed factor and \textit{Subject} as a random factor. \textit{Post hoc} pairwise comparisons were corrected with Holm's sequential Bonferroni procedure~\cite{holm1979simple} following significant omnibus tests.

We ran a paired-samples \textit{t}-test \cite{student1908probable} to compare text entry throughput before and after iterative customization, as throughput had a single value per participant per technique. We ran Wilcoxon signed-rank tests \cite{wilcoxon1945individual} to compare the task load and usability ratings before and after iterative customization. We also ran an LMM \cite{frederick1999fixed, littell1998statistical} to compare the transcription WPM and binomial generalized linear mixed models (GLMM)~\cite{brooks2017glmmtmb} for UER, CER, TER before and after iterative customization, using \textit{Technique} (\textit{Custom-Initial} \textit{vs.} \textit{Custom-Iterated}) as a fixed factor and \textit{Subject} and \textit{Phrase} as random factors. 

Given the small sample size (10 participants, 765 trials), our analyses have limited statistical power for detecting population-level effects. Moreover, participants exhibited heterogeneity in their text entry behavior, performance, and preferences, which may further obscure effects that generalize across participants. Therefore, in addition to inferential statistical analyses, we report descriptive statistics characterizing \textit{AdaptType}'s performance, task load, usability, and the coverage of ability-based recommendations. We also examine participants' customization choices and individual differences on a case-by-case basis to capture insights that may not be reflected in aggregate results.

%% file: 06-results.tex
\begin{figure*}[t]
    \vspace{-0.5pc}
    \includegraphics[width=0.85\textwidth]{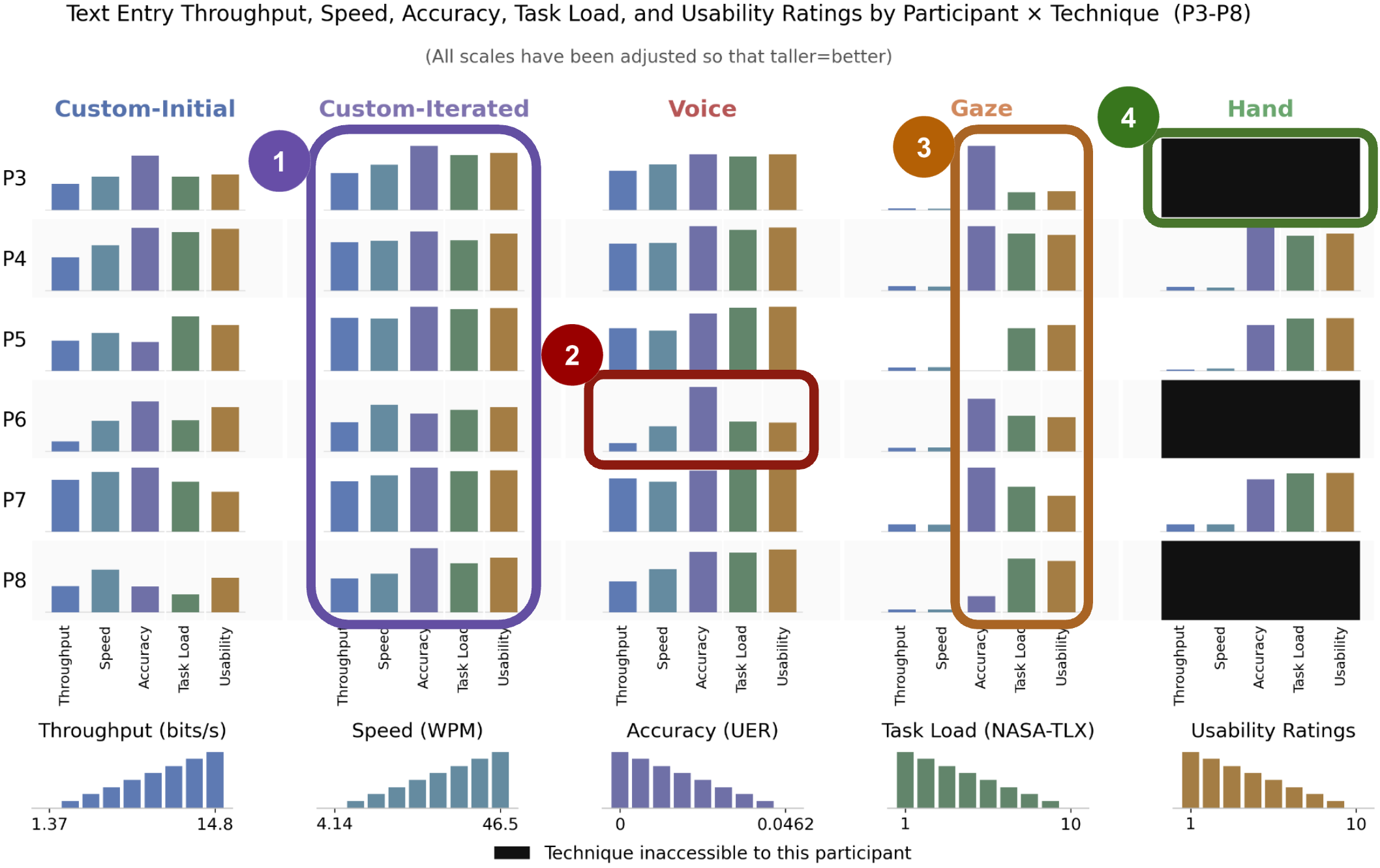}
    \vspace{-0.5pc}
    \caption{Participant-level comparison of five techniques in performance (throughput, WPM, UER), average task load, and average usability measures, among the six participants who completed both the \textit{initial} and the \textit{iterated} customization. The scales have been adjusted so that ``taller'' bars always indicate ``better'' experiences. A black cell indicates that technique is completely inaccessible to the participant. Participants' experiences with different modalities were highly varied: (1) the \textit{Custom-Iterated} technique exhibited consistently high performance, low task load, and high usability; (2) \textit{Voice} worked well for many participants, yet still failed for some (e.g., P6); (3) \textit{Gaze} was highly varied across participants; and (4) \textit{Hand} was entirely inaccessible to about half of the participants.}
    \vspace{-0.5pc}
    \Description{Participant-level comparison of five text entry techniques for participants P3--P8. Columns represent Custom-Initial, Custom-Iterated, Voice, Gaze, and Hand, and rows represent individual participants. Each participant-by-technique cell contains five bars for throughput, speed, accuracy, task load, and usability, with all measures normalized so that taller bars indicate better outcomes. Reference scales for the five measures are shown along the bottom. Black cells indicate that a technique was inaccessible to a participant. Four numbered outlines highlight the Custom-Iterated column, P6's Voice results, the Gaze column, and inaccessible Hand conditions, corresponding to observations discussed in the figure caption.}
    \label{fig:individual-differences}
\end{figure*}

\section{Results}

In this section, we report our user study results and how they address our two research questions: 
\begin{itemize}
    \item \textbf{RQ1.} Do customized multimodal text entry methods using \textit{AdaptType} make VR text entry more accessible for people with upper-body motor impairments? If so, in which respects?
    \item \textbf{RQ2.} What approaches can effectively support customization of VR text entry for people with upper-body motor impairments? 
\end{itemize}

\subsection{RQ1. Does AdaptType Make VR Text Entry More Accessible?}

\subsubsection{\textbf{Finding \#1: \textit{AdaptType} yielded high performance, low task load, and high usability across diverse motor abilities}}
\label{sec:finding-1}

Among the six participants who completed the iterated customization phase (Part~3, Section~\ref{sec:procedure}), they achieved an average text entry speed of 36.26 WPM ($SD=11.36$) and uncorrected error rate of 0.36\% ($SD=2.52\%$) using the \textit{Custom-Iterated} technique, and it exhibited the highest throughput (10.12 bits/s, $SD=2.06$) among all techniques (compared to 9.19 bits/s for \textit{Voice}, 7.45 bits/s for \textit{Custom-Initial}, 2.06 bits/s for \textit{Hand}, and 2.02 bits/s for \textit{Gaze}). Participants found their \textit{Custom-Iterated} technique to exhibit low physical demand ($M=2.2, SD=1.3$), low mental demand ($M=2.2, SD=1.0$), high performance ($M=3.2, SD=1.5$), low effort ($M=2.8, SD=1.3$), and low frustration ($M=2.5, SD=1.1$). Participants also found their \textit{Custom-Iterated} technique easy to use ($M=2.2, SD=1.2$), comfortable ($M=2.0, SD=1.1$), and causing little fatigue ($M=2.7, SD=0.8$). Participants indicated very high interest in future use ($M=1.7, SD=1.0$) of this technique. Participants showed high satisfaction with their iterated customization, regarding their customized techniques as ``\textit{this works great}'' (P3) and ``\textit{that's the one}'' (P7). P8 also commented that ``\textit{this is really easy, I'm super excited about it}.''

\begin{figure*}[t]
    \centering
    \vspace{-0.5pc}
    \includegraphics[width=\textwidth]{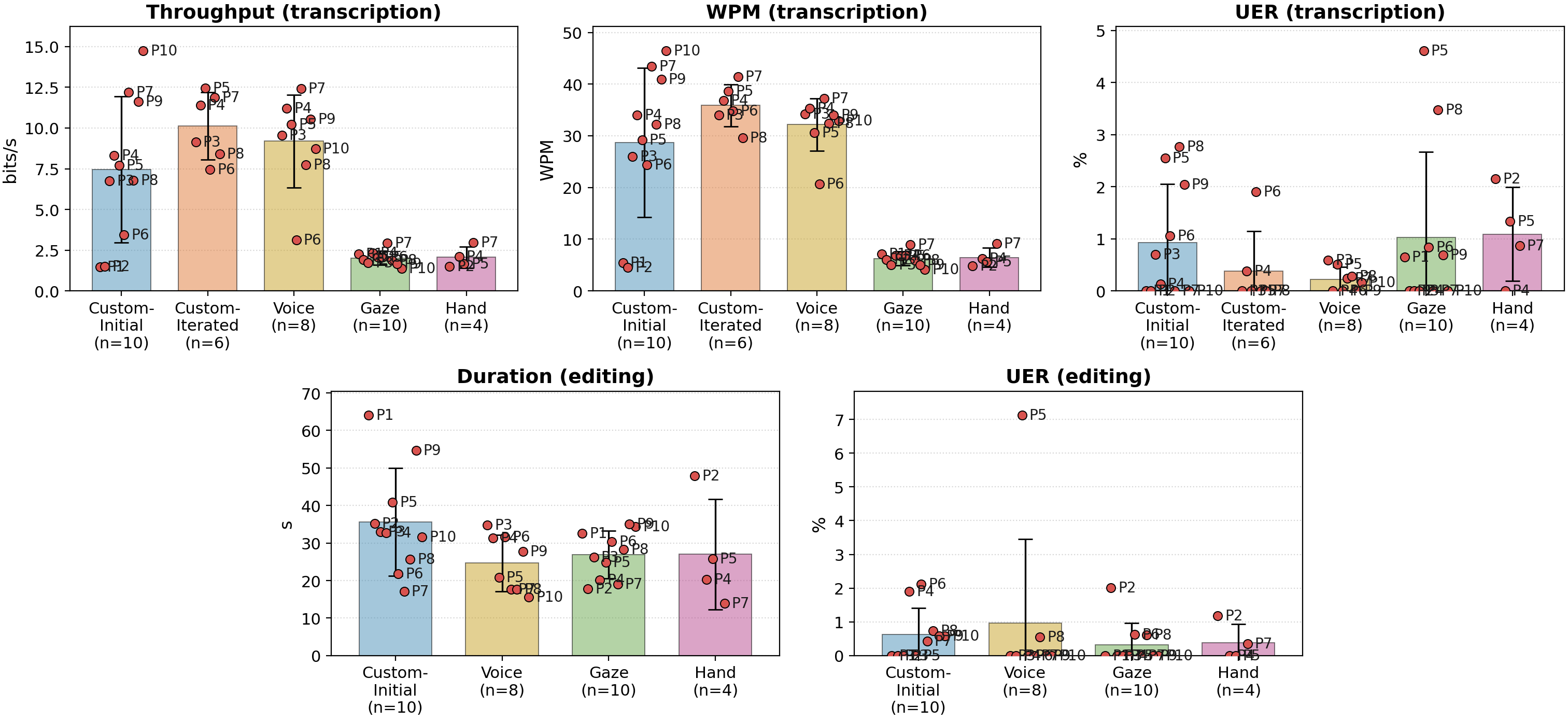}
    \vspace{-1.5pc}
    \caption{Performance measures of the five techniques in the transcription and editing tasks. Note that \textit{Custom-Iterated} is included in the figure to show the overall performance of different techniques, but was excluded from the comparison analysis (Section~\ref{sec:finding-2}) as it was not counterbalanced for \textit{Order}. Overall, \textit{Custom} (both \textit{Initial} and \textit{Iterated}) and \textit{Voice} techniques exhibited high speed and throughput, while \textit{Gaze} and \textit{Hand} were relatively slow and inefficient. In the \textbf{\textit{transcription}} tasks when participants entered sentences from scratch, \textit{Voice} exhibited the lowest uncorrected error rate. However, in the \textbf{\textit{editing}} tasks when participants corrected existing errors in sentences, \textit{Voice} exhibited the highest error rate, and became especially inaccurate for some participants (e.g., P5), while other techniques remained low in error rate. Put together, customized multimodal techniques offer both the benefit of fast input comparable to \textit{Voice} and high accuracy in error correction comparable to \textit{Gaze} or \textit{Hand}.} 
    \Description{Five bar charts compare the text entry techniques on transcription throughput, transcription speed, transcription uncorrected error rate, editing duration, and editing uncorrected error rate, with individual participant observations overlaid. Custom-Initial, Custom-Iterated, and Voice show substantially higher transcription speed and throughput than Gaze and Hand. Voice has the lowest transcription error rate but the highest editing error rate and substantial participant variation, whereas the customized techniques combine relatively fast entry with low editing error rates.}
    \vspace{-0.5pc}
    \label{fig:speed-and-accuracy}
\end{figure*}

\subsubsection{\textbf{Finding \#2: Customized, multimodal techniques exhibited text entry speed comparable to \textit{Voice} and higher than \textit{Gaze} and \textit{Hand}}}
\label{sec:finding-2}

With regard to text entry \textbf{speed}, both \textit{Custom-Initial} and \textit{Voice} had relatively high text entry speed in WPM ($M=35.03$, $SD=16.78$ and $M=33.43$, $SD=9.24$, respectively) in the transcription task, compared to \textit{Gaze} ($M=6.54$, $SD=1.63$) and \textit{Hand} ($M=6.82$, $SD=2.09$). \textit{Technique} had a significant effect on \textit{WPM} ($F_{3,444.3} = 120.44, p<.001$). The \textit{Custom-Initial} technique had higher WPM than \textit{Gaze} ($t_{446}=16.12, p<.001$) and \textit{Hand} ($t_{441}=11.10, p<.001$); the \textit{Voice} technique had higher WPM than \textit{Gaze} ($t_{442}=14.83, p<.001$) and \textit{Hand} ($t_{442}=10.32, p<.001$). We found no significant differences in WPM between \textit{Custom-Initial} and \textit{Voice} ($t_{448}=1.83, p=.135$), or between \textit{Gaze} and \textit{Hand} ($t_{441}=0.69, p=.489$). In the editing task, \textit{Technique} had a significant effect on \textit{Duration} ($F_{3,166.3}=2.85, p=.039$): \textit{Custom-Initial} took longer than \textit{Voice} ($M=33.7$s \textit{vs.} $24.4$s; $t_{165}=2.74, p=.041$), with no other pairwise difference reaching significance.

\begin{figure*}[h]
    \vspace{-0.5pc}
    \includegraphics[width=0.7\textwidth]{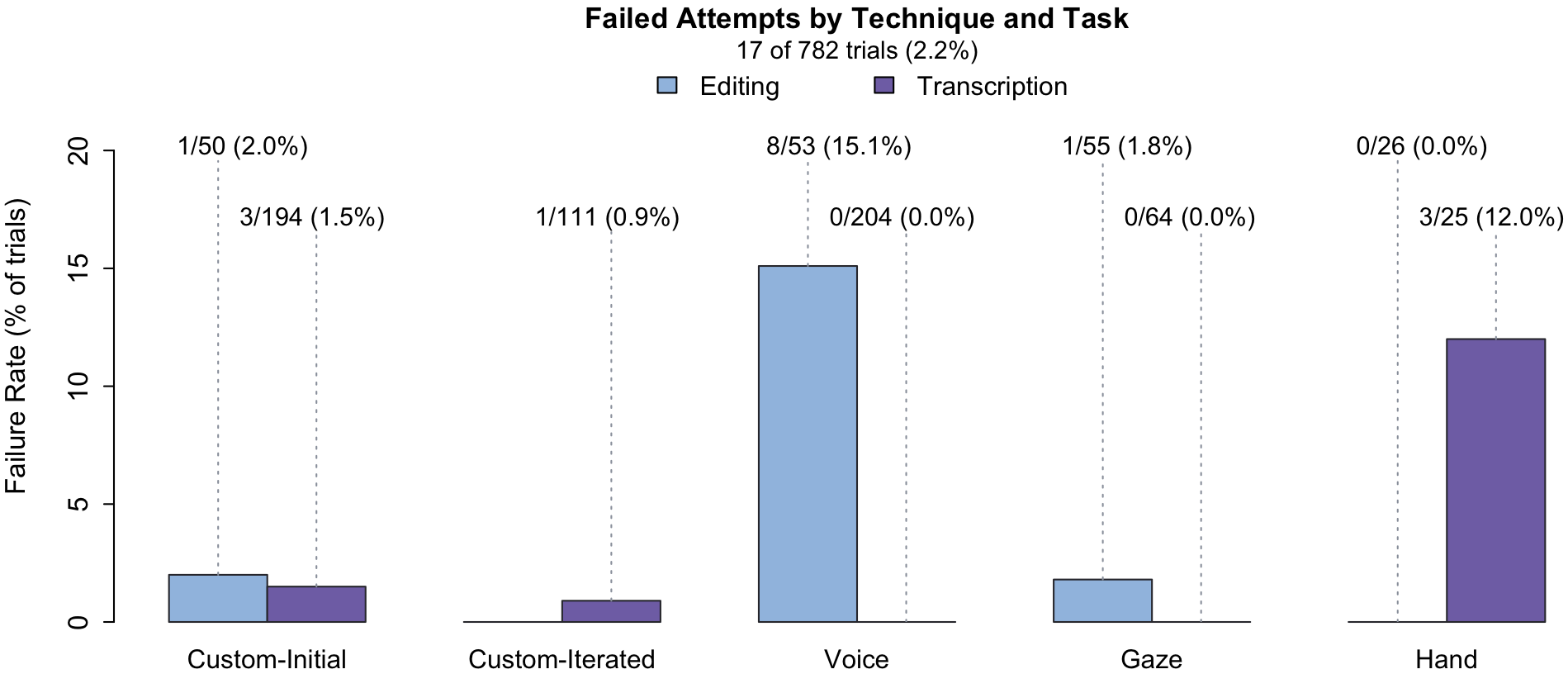}
    \vspace{-0.5pc}
    \caption{The rate of failed attempts for each technique by task. 
    \textit{Voice} had a zero failure rate with transcription, yet had a high failure rate (15.1\%) with editing, when participants were instructed to correct errors in an existing sentence. \textit{Hand} had a zero failure rate with editing, yet had a high failure rate (12.0\%) with transcription, when participants were instructed to enter a full sentence from scratch. Note that only transcription data is available for the \textit{Custom-Iterated} technique. Among the rest of the four techniques, \textit{Custom-Initial} and \textit{Gaze} were the only techniques exhibiting consistently low failure rates ($\leq2\%$) across editing and transcription tasks.}
    \vspace{-0.5pc}
    \Description{Grouped bar chart showing the percentage of failed attempts for each technique in editing and transcription tasks. Voice has no failed transcription trials but the highest editing failure rate at 15.1\%; Hand has no failed editing trials but the highest transcription failure rate at 12.0\%. Custom-Initial and Gaze maintain failure rates at or below 2\% across both tasks; Custom-Iterated is shown only for transcription.}
    \label{fig:failed-attempts}
\end{figure*}

\begin{table*}[h]
\centering
\resizebox{\textwidth}{!}{%
\begin{tabular}[t]{|l|c|c|c|c|}
\hline
 & \multicolumn{3}{c|}{\textbf{Transcription}} & \multicolumn{1}{c|}{\textbf{Editing}} \\ \cline{2-5}
 & \textbf{UER} & \textbf{CER} & \textbf{TER} & \textbf{UER} \\ \cline{2-5}
 & \textit{\small M (SD)} & \textit{\small M (SD)} & \textit{\small M (SD)} & \textit{\small M (SD)} \\ \hline
 \textit{Custom-Initial} & 0.92 (1.13) & 7.44 (7.19) & 8.37 (7.06) & 0.63 (0.78) \\ \hline
\textit{Custom-Iterated} & 0.38 (0.76) & 6.12 (6.57) & 6.50 (6.49) & n/a \\ \hline
\textit{Voice} & 0.22 (0.23) & 4.42 (5.47) & 4.64 (5.39) & 0.96 (2.50) \\ \hline
\textit{Gaze} & 1.03 (1.65) & 3.83 (3.76) & 4.85 (4.25) & 0.32 (0.64) \\ \hline
\textit{Hand} & 1.09 (0.90) & 3.04 (1.28) & 4.12 (1.99) & 0.38 (0.55) \\ \hline
\end{tabular}
\hspace{1.5em}
\begin{tabular}[t]{|l|c|c|c|}
\hline
 & \multicolumn{3}{c|}{\textbf{Transcription}} \\ \cline{2-4}
 & \textbf{UER} & \textbf{CER} & \textbf{TER} \\ \hline
\textit{Custom-Initial--Voice} & 3.07 * & 3.49 ** & 4.54 *** \\ \hline
\textit{Custom-Initial--Gaze} & --- & --- & --- \\ \hline
\textit{Custom-Initial--Hand} & --- & -2.18 $\cdot$ & --- \\ \hline
\textit{Voice--Gaze} & -2.43 $\cdot$ & -4.04 *** & -4.55 *** \\ \hline
\textit{Voice--Hand} & -3.09 * & -4.35 *** & -4.36 *** \\ \hline
\textit{Gaze--Hand} & --- & --- & --- \\ \hline
\end{tabular}
}
\caption{Left: error rates by technique, as participant means in percent. Right: $z$-statistics for significant or marginal pairwise comparisons. Measures with no significant or marginal differences are omitted from the right-hand block. \textit{Custom-Iterated} is reported descriptively but excluded from the comparisons, as it always ran last and was not counterbalanced for \textit{Order}. CER and TER do not apply to the editing task, where participants corrected pre-seeded errors rather than entering text from scratch. Significance indicators: $\cdot\, p<.10, \text{*} p<.05, \text{**} p<.01, \text{***} p<.001$.}
\Description{Error-rate results and statistical comparisons by technique. The left portion reports participant-mean uncorrected, corrected, and total error rates for transcription and uncorrected error rate for editing. Voice has the lowest transcription uncorrected error rate at 0.22\% but the highest editing uncorrected error rate at 0.96\%; Gaze and Hand have lower editing error rates. The right portion reports significant or marginal pairwise z-statistics for transcription error measures, with the strongest differences generally involving Voice.}
\vspace{-1pc}
\label{tab:error-rates}
\end{table*}

\subsubsection{\textbf{Finding \#3: \textit{Voice} excelled in transcription accuracy and task load, yet struggled with error correction, which multimodality remedied}}
\label{sec:finding-3}

With regard to text entry \textbf{accuracy}, \textit{Technique} had a significant effect on \textit{UER} ($\chi^2_3=12.66, p=.005$), \textit{CER} ($\chi^2_3=24.48, p<.001$) and \textit{TER} ($\chi^2_3=29.80, p<.001$) in the transcription task---while all techniques exhibited low UER ($<1.5\%$), \textit{Voice} had the lowest UER of all at 0.22\% ($SD=0.23$). For the editing task, we omit statistical analysis with UER, as only 17 of the 174 trials (9.8\%) had a non-zero UER. Descriptively, \textit{Voice} had the highest editing UER ($M=0.96\%, SD=2.50\%$), compared to \textit{Custom-Initial} ($M=0.63\%, SD=0.78\%$), \textit{Gaze} ($M=0.32\%, SD=0.64\%$) and \textit{Hand} ($M=0.38\%, SD=0.55\%$). \textit{Voice} also had the highest variance in UER across participants, with the highest being 7.1\% (P5), while all other techniques' highest participant-average UER remained below 3\% (Figure~\ref{fig:speed-and-accuracy}). 

Notably, the accuracy analysis was conducted based on \textit{successful} trials after removing failed attempts. With regard to \textbf{failed attempts}, however, \textit{Voice} had the highest failure rate in the editing task (15.1\%) among all techniques, compared to 2.0\% for \textit{Custom-Initial}, 1.8\% for \textit{Gaze}, 0.0\% for \textit{Hand}. In the transcription task, \textit{Hand} had the highest failure rate (12.0\%) among all techniques, compared to 1.5\% for \textit{Custom-Initial}, 0.9\% for \textit{Custom-Iterated}, and 0.0\% for \textit{Voice}. Among the four techniques for which both editing and transcription data is available, only \textit{Custom-Initial} and \textit{Gaze} exhibited consistently low failure rate within 2\% across editing and transcription tasks. 

\begin{table*}[]
\centering
\vspace{-0.5pc}
\resizebox{\textwidth}{!}{%
\begin{tabular}{|l|c|c|c|c|c|c||c|c|c|c|c|}
\hline 
 & \multicolumn{6}{c||}{\textbf{Task Load}} & \multicolumn{5}{c|}{\textbf{Usability}} \\ \cline{2-12}
 & \textbf{\begin{tabular}[c]{@{}c@{}}Mental\\ Demand\end{tabular}} & \textbf{\begin{tabular}[c]{@{}c@{}}Physical\\ Demand\end{tabular}} & \textbf{\begin{tabular}[c]{@{}c@{}}Temporal\\ Demand\end{tabular}} & \textbf{\begin{tabular}[c]{@{}c@{}}Perfor-\\ mance\end{tabular}} & \textbf{\begin{tabular}[c]{@{}c@{}}Effort\end{tabular}} & \textbf{\begin{tabular}[c]{@{}c@{}}Frustration\end{tabular}} & \textbf{\begin{tabular}[c]{@{}c@{}}Ease of\\ Use\end{tabular}} & \textbf{\begin{tabular}[c]{@{}c@{}}Ease of\\ Learning\end{tabular}} & \textbf{\begin{tabular}[c]{@{}c@{}}Comfort\end{tabular}} & \textbf{\begin{tabular}[c]{@{}c@{}}Fatigue\end{tabular}} & \textbf{\begin{tabular}[c]{@{}c@{}}Future\\ Use\end{tabular}} \\ \hline
\multicolumn{12}{|l|}{\textbf{Transcription}} \\ \hline
\textit{Custom-Initial--Voice} & 2.70 * & 2.78 * & --- & 2.54 $\cdot$ & 3.29 ** & 3.22 ** & 3.23 ** & --- & 3.12 * & 2.61 $\cdot$ & 2.50 $\cdot$ \\ \hline
\textit{Voice--Gaze} & -2.58 * & -2.50 $\cdot$ & --- & --- & -3.10 ** & -2.85 * & -2.68 * & --- & -2.67 * & -2.45 $\cdot$ & -2.50 $\cdot$ \\ \hline
\textit{Voice--Hand} & --- & -2.43 $\cdot$ & --- & --- & -2.26 $\cdot$ & --- & -2.30 $\cdot$ & --- & --- & --- & --- \\ \hline
\multicolumn{12}{|l|}{\textbf{Editing}} \\ \hline
\textit{Custom-Initial--Voice} & --- & --- & --- & --- & 2.40 $\cdot$ & 2.74 * & --- & --- & --- & --- & --- \\ \hline
\textit{Voice--Gaze} & --- & --- & --- & --- & --- & -2.63 * & --- & --- & --- & --- & --- \\ \hline
\end{tabular}
}
\vspace{0.5pc}
\caption{$Z$-statistics for significant or marginal pairwise comparisons on the NASA-TLX task load and usability items, from mixed ordinal logistic regressions with Holm correction within each item. A positive $Z$ means the first technique was rated less favorably. Technique pairs with no significant or marginal comparison on any item are omitted. \textit{Custom-Iterated} is excluded, as it always ran last, was not counterbalanced for \textit{Order}, and had no editing task. Significance indicators: $\cdot\, p<.10, \text{*} p<.05, \text{**} p<.01, \text{***} p<.001$.}
\Description{Significant or marginal pairwise comparisons among techniques for six NASA-TLX task-load measures and five usability measures, reported separately for transcription and editing. During transcription, numerous comparisons favor Voice over Custom-Initial, Gaze, or Hand, particularly for effort, frustration, ease of use, and comfort. During editing, differences are substantially fewer, with significant differences remaining primarily for frustration and a marginal difference for effort.}
\vspace{-1pc}
\label{tab:task-load-usability-ratings}
\end{table*}

With regard to \textbf{task load} and \textbf{usability}, \textit{Technique} had a significant effect on mental demand ($\chi^2_3=10.13, p=.017$), physical demand ($\chi^2_3=11.06, p=.011$), performance ($\chi^2_3=9.18, p=.027$), effort ($\chi^2_3=15.72, p=.001$), frustration ($\chi^2_3=14.92, p=.002$), ease of use ($\chi^2_3=14.17, p=.003$), comfort ($\chi^2_3=13.25, p=.004$), fatigue ($\chi^2_3=8.63, p=.035$), and future use ($\chi^2_3=10.25, p=.017$) in the transcription task, with a marginal effect on temporal demand ($\chi^2_3=7.15, p=.067$). In the editing task, \textit{Technique} had a significant effect on frustration ($\chi^2_3=10.10, p=.018$) and a marginal effect on effort ($\chi^2_3=6.75, p=.080$). All significant or marginal pairwise comparisons are shown in Table~\ref{tab:task-load-usability-ratings}. All pairwise comparisons favored \textit{Voice}, yet fewer ratings showed significant differences in \textit{editing} than \textit{transcription} in omnibus tests. Meanwhile, individual differences remained. For example, P6 rated \textit{Voice} to have the highest average task load ($M=5.83,SD=0.41$) and worst usability ($M=6.00,SD=0.00$) among all techniques. 

\subsection{RQ2. What approaches can effectively support customization of VR text entry?}

\begin{table*}[t]
\vspace{-0.5pc}
\centering
\resizebox{\textwidth}{!}{%
\begin{tabular}{@{}ccccccccccccc@{}}
\toprule
 & \multicolumn{3}{c}{\textit{\textbf{Initial} Customization Choices}}
 & \multicolumn{4}{c}{\textit{\textbf{Iterated} Customization Choices}}
 & \multicolumn{2}{c}{\textit{\textbf{\begin{tabular}[c]{@{}c@{}}Self-Reported\\ Fine Motor Function\end{tabular}}}}
 & \multicolumn{2}{c}{\textit{\textbf{\begin{tabular}[c]{@{}c@{}}Self-Reported\\ Gross Motor Function\end{tabular}}}}
 & \textit{\textbf{\begin{tabular}[c]{@{}c@{}}Self-Rep.\\ Voice\end{tabular}}} \\
 & \multicolumn{3}{c}{\textit{(Before Extended Usage)}}
 & \multicolumn{4}{c}{\textit{(After Extended Usage)}}
 & \multicolumn{2}{c}{\textit{(1=no difficulty, 5=unable)}}
 & \multicolumn{2}{c}{\textit{(1=no difficulty, 5=unable)}}
 & \textit{(1--5)} \\
\cmidrule(lr){2-4} \cmidrule(lr){5-8} \cmidrule(lr){9-10} \cmidrule(lr){11-12} \cmidrule(lr){13-13}
\textit{\textbf{ID}} &
  \textit{\textbf{\begin{tabular}[c]{@{}c@{}}Input\\ Modality\end{tabular}}} &
  \textit{\textbf{\begin{tabular}[c]{@{}c@{}}Auxiliary\\ Pointer\end{tabular}}} &
  \textit{\textbf{\begin{tabular}[c]{@{}c@{}}Selection\\ Trigger\end{tabular}}} &
  \textit{\textbf{\begin{tabular}[c]{@{}c@{}}Input\\ Modality\end{tabular}}} &
  \textit{\textbf{\begin{tabular}[c]{@{}c@{}}Auxiliary\\ Pointer\end{tabular}}} &
  \textit{\textbf{\begin{tabular}[c]{@{}c@{}}Selection\\ Trigger\end{tabular}}} &
  \textit{\textbf{\begin{tabular}[c]{@{}c@{}}Additional\\ Feature\end{tabular}}} &
  \textit{\textbf{Left}} & \textit{\textbf{Right}} &
  \textit{\textbf{Left}} & \textit{\textbf{Right}} & \\
\midrule
\textbf{P1}  & Voice      & Gaze       & Dwell & -- & -- & -- & -- & 5 & 5 & 5 & 5 & 3 \\ \midrule
\textbf{P2}  & Right Hand & --         & Left Pinch & -- & -- & -- & -- & 1 & 1 & 1 & 1 & 1 \\ \midrule
\textbf{P3}  & Voice & Right Hand & \begin{tabular}[c]{@{}c@{}}Left Hand\\ (curling left wrist\\ on armrest)\end{tabular} & Voice & Right Hand & \begin{tabular}[c]{@{}c@{}}Left Hand\\ (curling left wrist\\ on armrest)\end{tabular} & Voice command & 5 & 5 & 2 & 2 & 1 \\ \midrule
\textbf{P4}  & Voice & Gaze & \begin{tabular}[c]{@{}c@{}}Right Finger Tap\\ (on desktop)\end{tabular} & Voice & Gaze & \begin{tabular}[c]{@{}c@{}}Right Finger Tap\\ (on desktop)\end{tabular} & Voice command & 2 & 3 & 3 & 2 & 2 \\ \midrule
\textbf{P5}  & Voice & Gaze & \begin{tabular}[c]{@{}c@{}}Right Finger Tap\\ (on desktop)\end{tabular} & Voice & Right Hand & Right Pinch & Voice command & 5 & 1 & 1 & 1 & 1 \\ \midrule
\textbf{P6}  & Voice & Gaze & \begin{tabular}[c]{@{}c@{}}Left Hand Tap\\ (on armrest)\end{tabular} & Voice & Gaze & \begin{tabular}[c]{@{}c@{}}Left Hand Tap\\ (on armrest)\end{tabular} & Voice command & 4 & 4 & 4 & 5 & 3 \\ \midrule
\textbf{P7}  & Voice & Gaze & \begin{tabular}[c]{@{}c@{}}Right Finger Tap\\ (on desktop)\end{tabular} & Voice & Gaze & Dwell & Voice command & 1 & 1 & 3 & 3 & 1 \\ \midrule
\textbf{P8}  & Voice & Head & \begin{tabular}[c]{@{}c@{}}Left Hand Tap\\ (on desktop)\end{tabular} & Voice & Gaze & Dwell & Voice command & 4 & 5 & 1 & 1 & 1 \\ \midrule
\textbf{P9}  & Voice & Gaze & \begin{tabular}[c]{@{}c@{}}Left Hand Tap\\ (on desktop)\end{tabular} & -- & -- & -- & -- & 4 & 4 & 4 & 5 & 1 \\ \midrule
\textbf{P10} & Voice & Gaze & Dwell & -- & -- & -- & -- & 4 & 2 & 3 & 1 & 1 \\
\bottomrule
\end{tabular}%
}
\vspace{0.5pc}
\caption{Customization choices and self-reported motor and voice function for each participant. Overall, participants selected customization options that aligned with their self-reported motor abilities. For example, P3 retained much better gross motor function than fine motor function on both sides, thus selected \textit{right hand} and \textit{left hand wrist curl} as the pointing and trigger gestures, which required no fine motor function; P7 retained much better fine motor function than gross motor function, thus selected the \textit{right finger desktop tap} trigger gesture and \textit{gaze dwell}, which required no gross motor function; P5 only had impaired fine motor function on the left side, thus selected the \textit{right hand} as the pointer and \textit{right finger tap} and \textit{right pinch} as the triggers.}
\Description{Participant-level initial and iterated customization choices alongside self-reported left and right fine motor function, gross motor function, and voice function. The table shows substantial diversity in selected input modalities, pointers, and triggers, while choices generally align with participants' reported abilities. Participants with severe bilateral fine motor impairment select gaze dwell or whole-hand/arm triggers, whereas participants retaining fine motor function commonly select finger-based triggers; most participants use voice as their primary input modality.}
\vspace{-1.5pc}
\label{tab:customization-choices}
\end{table*}

\subsubsection{\textbf{Finding \#4: AdaptType's ability-based recommendations effectively scaffolded selection of text entry methods that matched participants' upper-body motor abilities}}
\label{sec:finding-4}

Participants selected a wide range of different input modalities, pointing methods, and selection triggers (Table~\ref{tab:customization-choices}). Nine of the 10 participants selected voice as their primary input modality, and one selected right hand. Seven participants selected gaze as the auxiliary pointer, one selected right hand, and one selected the head pointer. Selection triggers were the most varied: finger tapping on the desktop (3), hand tapping on the desktop (2) or on the armrest (1), gaze dwelling (2), wrist curling up on the armrest (1), and pinch (1). In the iterated customization choices, all six participants selected voice as their primary input modality; four participants selected gaze as the auxiliary pointer, and two selected the right hand. Participants also chose a variety of selection triggers, including hand tapping on the desktop (1) or on the armrest (1), finger tapping on the desktop (1), wrist curling on the armrest (1), gaze dwelling (2), and pinch (1).

Participants selected customization options that aligned with their motor abilities (Table~\ref{tab:customization-choices}). All five participants with severe bilateral fine motor impairment (self-reported 4 or 5 on both sides) selected either gaze dwelling or a whole-hand/arm selection trigger, requiring no fine motor function. All three participants who retained unimpaired fine motor function on at least one side (self-reported 1) selected a finger-based trigger on that side. P1 reported no fine or gross motor function on either side (5 across all four ratings), and selected a fully limb-free configuration using voice and gaze. P10 was the only exception to this pattern, selecting the same limb-free configuration despite retaining moderate gross and fine motor function on the right side, suggesting that limb-free configurations could be selected by preference as well as by necessity. Overall, participant selected customization options aligned with their motor abilities. This alignment additionally addresses \textbf{RQ1}, showing that \textit{AdaptType} enabled users to select input modalities and triggers that suited their abilities and needs across diverse motor abilities. 

\begin{figure}[h]
    \vspace{-0.5pc}
    \includegraphics[width=\linewidth]{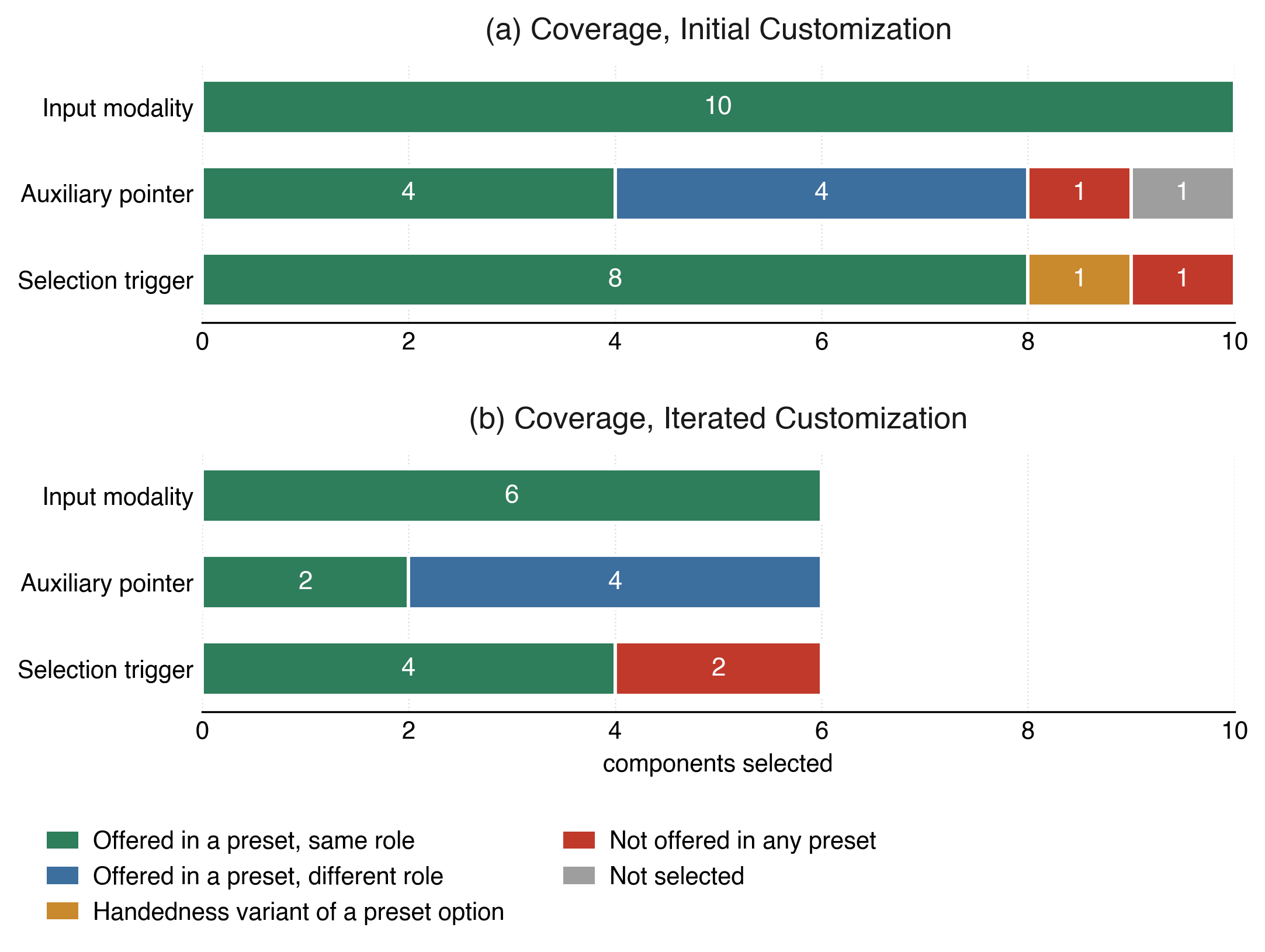}
    \vspace{-1pc}
    \caption{Coverage of users' customization choices in recommended presets for (a) \textit{initial} customization and (b) \textit{iterated} customization. Green bars represent participant choices offered in a preset and with the same role. Blue bars represent choices offered in a preset with a different role, e.g., when \textit{gaze} was offered as an \textit{input modality}, yet a participant selected it as the \textit{auxiliary pointer}. Yellow bars represent handedness variant of a preset option, e.g., when \textit{right pinch} was offered in a preset, yet a participant selected \textit{left pinch}. Red bars represent selections not offered in any preset, and gray bars represent cases when the auxiliary pointer was not selected (not applicable). All participants selected input modalities offered in the recommendations during both \textit{initial} and \textit{iterated} customization. For the auxiliary pointer, most participants' choices were offered in the recommendations, but some with a different role. For the selection trigger, most participants' choices were offered in the recommendations, while some updated their choices to triggers not offered in any preset in the \textit{iterated} customization.}
    \vspace{-1pc}
    \Description{Two stacked horizontal bar charts show how well participants' selected input modalities, auxiliary pointers, and selection triggers were covered by AdaptType's recommended presets, separately for initial and iterated customization. During initial customization, recommended presets covered all selected input modalities and 80\% of selected auxiliary pointers and triggers. During iterated customization, presets covered all selected input modalities and auxiliary pointers and four of six selected triggers. Colors distinguish whether the selected component was offered in the same role, offered in a different role, represented by a handedness variant, not offered, or not selected.}
    \label{fig:coverage-analysis-chart-only}
\end{figure}

The ability-based recommendations offered by \textit{AdaptType} (Section~\ref{sec:ability-based-recommendations-of-customization-settings}) also aligned with participants' customization choices (Figure~\ref{fig:coverage-analysis-chart-only}). Among the 10 participants, all selected input modalities offered in the three presets recommended (100\%), eight selected auxiliary pointers offered in the recommendations (80\%), and eight selected triggers offered in the recommendations (80\%). Among the six participants who completed the iterated customization, all six selected input modalities (100\%) and auxiliary pointers (100\%) offered in the recommendations, and four selected triggers offered in the recommendations (66.7\%). The alignment between the recommendations and participants' customization choices indicates the effectiveness of ability-based recommendations in scaffolding customization. 

\subsubsection{\textbf{Finding \#5: Iterative customization showed trends of improvement in text entry performance, task load, and usability}}
\label{sec:finding-5}

After iterative customization, the average text entry throughput increased from 7.53 bits/s ($SD=2.83$) to 10.12 bits/s ($SD=2.06$), with a 34.4\% improvement from \textit{Custom-Initial} to \textit{Custom-Iterated}. The paired-samples \textit{t}-test showed a significant difference ($t_5 = 3.51, p = .017$). The average text entry speed increased slightly from 31.54 WPM ($SD=6.88$) to 35.88 WPM ($SD=4.08$), with a 13.7\% improvement from \textit{Custom-Initial} to \textit{Custom-Iterated}. We found a marginal effect of \textit{Technique} on text entry speed (WPM) ($F_{1,203.6}=3.57, p=.060$), a significant effect on UER ($\chi^2_1=6.28, p=.012$), which decreased from 0.27\% to 0.05\%, a marginal effect on TER ($\chi^2_1=3.63, p=.057$), and no effect on CER ($\chi^2_1=2.07, p=.150$). 

\begin{figure*}[t]
    \vspace{-0.5pc}
    \includegraphics[width=0.7\textwidth]{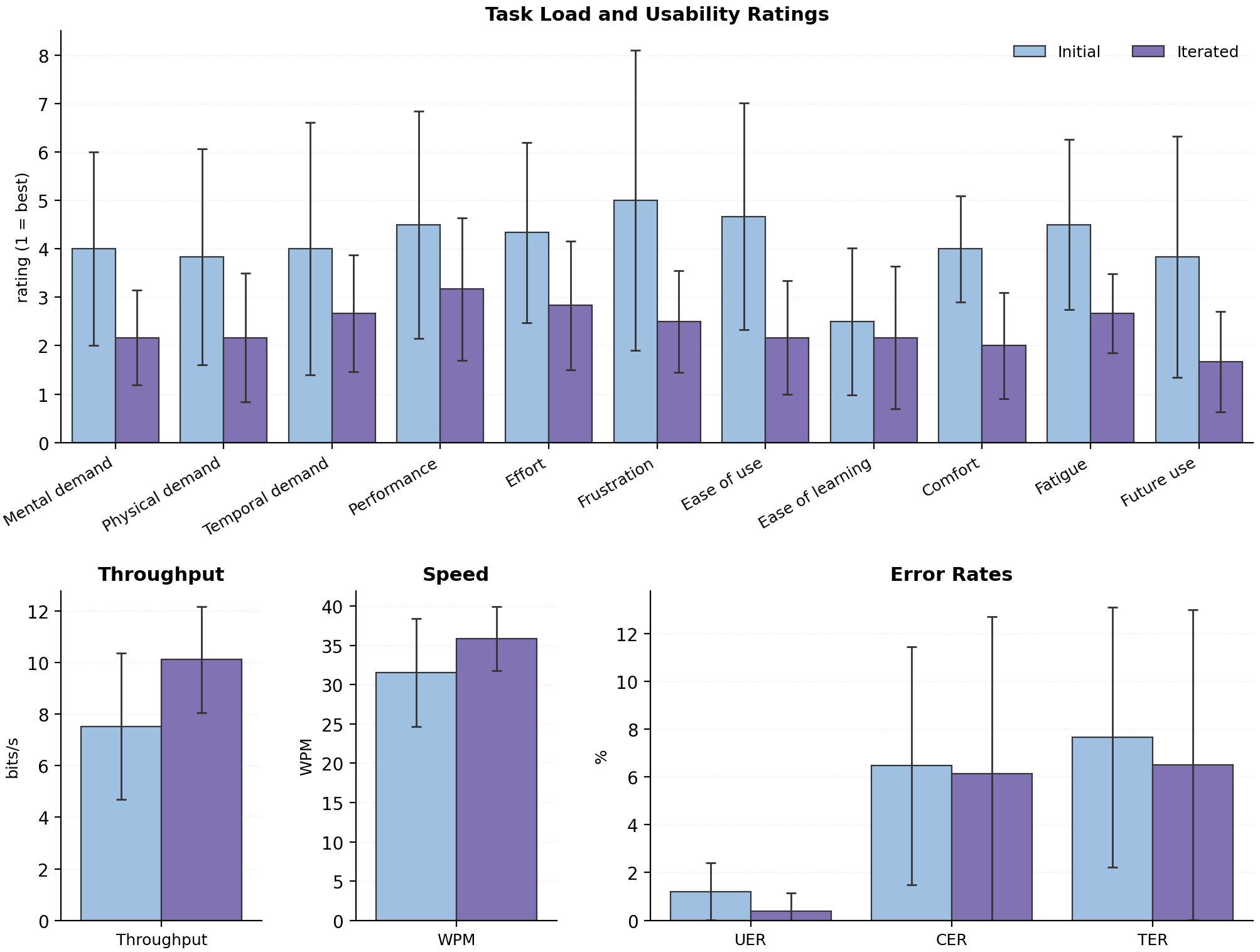}
    \vspace{-0.5pc}
    \caption{$M\pm SD$ of performance measures, task load, and usability ratings of the \textit{initial} \textit{vs.} \textit{iterated} customization ($n=6$).}
    \vspace{-1pc}
    \Description{Grouped bar charts compare initial and iterated customization for the six participants who completed both phases. The upper chart shows NASA-TLX and usability ratings, with iterated customization generally receiving lower, more favorable ratings across measures. The lower charts show higher throughput and text entry speed after iteration, lower uncorrected error rate, and modest reductions in corrected and total error rates. Error bars show mean plus or minus one standard deviation.}
    \label{fig:iterative-customization}
\end{figure*}

Among task load and usability ratings, marginal significance was detected in Wilcoxon signed-rank tests with mental demand ($z=-2.07, p=.053$), physical demand ($z=-1.80, p=.090$), and comfort ($z=-2.00, p=.072$). No other significant or marginal differences were found. Although only marginal significances were detected, descriptive statistics showed consistent trends of improvement after iterative customization (Figure~\ref{fig:iterative-customization}). Participants also reported having better understanding of their ideal customization after extended usage and trial-and-error with different input modalities over tasks. For example, P7 switched from hand-based trigger to dwell trigger, as they ``\textit{liked the flow once [they were] familiar with it}''; and P5 switched from gaze-based pointing to hand-based pointing, as they had enough gross motor function to manipulate the hand pointer, yet gaze tracking ``\textit{didn't work well}'' for them.

%% file: 07-discussion.tex
\section{Discussion}

In this section, we discuss findings from our case studies, implications for future designs, and limitations of this work. 

\subsection{Accommodating a Range of Motor Abilities}

A central motivation for \textit{AdaptType} is that motor impairments are \textit{heterogeneous}---some users retain strong gross motor ability but limited finger dexterity, while others may experience the reverse~\cite{yamagami2023people}. Our 10 participants had a wide range of upper-body motor disabilities that affected different aspects of their gross and fine motor abilities, from spinal cord injury and amputation to neuropathy and muscular dystrophy. Our study results illustrate how the customization framework accommodated this diversity: participants with severe bilateral fine motor impairment (e.g., P1, P3, P6, P8) selected gaze- or hand/arm-based selection triggers, while participants who retained unimpaired fine motor function on at least one side selected a finger-based technique (e.g., P2, P5, P7) (Section~\ref{sec:finding-4}). Across participants with different motor abilities, \textit{AdaptType} yielded high performance (36.26 WPM, 0.36\% uncorrected error rate, 10.12 bits/s), low task load ($M=2.58, SD=1.08$), and high usability ($M=2.13, SD=0.93$), which demonstrates the system's capability in supporting accessible VR text entry for people with varying upper-body motor impairments (Section~\ref{sec:finding-1}).

Beyond findings that characterize holistic behavior patterns and preferences of all participants, individual differences remained, which further highlights the heterogeneity of the population and the need for customization---for example, P10 found \textit{Gaze} highly usable, while P3 experienced extremely high task load and low usability with it; P9 rated \textit{Voice} to be the best in task load and usability among all techniques, yet P6 had the worst experience with it (Figure~\ref{fig:individual-differences}). Such variations indicate that there is no single best solution in VR text entry for people with upper-body motor impairments. 

The central contribution of \textit{AdaptType} is not that it \textit{outperforms} every single baseline in every respect, but rather that it \textit{enables} users with diverse motor abilities to configure input methods in a way that best suits their abilities. 

\subsection{Why Not Just Use Voice Input?}

Admittedly, voice is a powerful modality in text entry for people with upper-body motor impairments~\cite{li2022freedom, rosenblatt2018vocal, okafor2023voice, dalla2018multi}. Its popularity and advantages were also reflected in our data: voice dictation had been widely adopted in daily usage among the participants ($M=3.3, SD=1.5$ in self-reported expertise, on a 1--5 scale), and nine out of 10 participants selected voice as their main input modality during customization, which may have partly contributed to the high text entry speed of the customized techniques. Our analyses also showed that \textit{Voice} overall exhibited high speed, low transcription accuracy, low task load, and high usability (Section~\ref{sec:finding-2} and~\ref{sec:finding-3}). 

However, voice-only techniques can be limited in some use cases, and prior work has highlighted the need for multimodal alternatives beyond single-modality input for people with upper-body motor impairments~\cite{li2022freedom}. Here, we discuss limitations of voice input that \textit{AdaptType} can remedy: 

\paragraph{\textbf{Voice does not work for everyone}} Despite the general trends that favored \textit{Voice} in our statistical analyses, pronounced individual differences remained. For example, P6 had facioscapulohumeral muscular dystrophy and self-reported having impaired voice (3 on a scale of 1--5). Accordingly, P6 rated the voice-only technique to have the highest task load ($M=5.83, SD=0.41$) and the lowest usability ($M=6.00, SD=0.00$) among all techniques and among all participants. Such experiences could occur for people with similar voice impairments, for example, due to age, dysarthria~\cite{malu2018exploring}, or other muscular or neurological conditions. 

\paragraph{\textbf{Voice is not as effective in error correction as in transcription}} 
Through our \textit{editing} task, we aimed to compare customized multimodal techniques against single-modality techniques when errors occurred. We showed that \textit{Voice} was highly effective in \textit{transcription} tasks; however, when it came to \textit{editing}, it had a notably high failure rate (15.1\%) and relatively high variation in uncorrected error rate among successful trials ($SD=2.5\%$, $\text{max}=7.1\%$). Meanwhile, although \textit{Voice} excelled in most \textit{transcription} task load and usability ratings, this advantage lessened when it came to \textit{editing} (Table~\ref{tab:task-load-usability-ratings}). 

\paragraph{\textbf{Multimodal techniques improve user agency compared to using voice only}} In our study, participants also indicated interest in using pointer-based techniques aside from voice input due to the capabilities of making fine edits. For example, P3 mentioned that ``there are still scenarios where further adjustments would be needed,'' and P4 liked the multimodal pointer-based technique because they ``feel like [they] have better control.'' 

\paragraph{\textbf{Voice may not be ideal for some usage scenarios}} Prior research has suggested that voice input may lead to privacy and security issues when used in public~\cite{easwara2015privacy, li2023security}. Aside from these concerns, participants also mentioned scenarios in which voice dictation had failed for them, for example, ``when I lost my voice'' (P3) or when ``listening to podcast'' (P7). Participants believed that multimodal text entry could be helpful in these scenarios when voice input may not be ideal. 

While we highlight these limitations of voice, this paper does not argue that pointer-based techniques should \textit{replace} voice input for people with upper-body motor impairments. Rather, customized multimodal techniques \textit{complement} voice dictation in cases where voice may fail or be less preferred---by supporting error correction, improving user agency, and offering flexibility in different usage scenarios. 

\begin{table*}[t]
\vspace{-0.7pc}
\centering
\small
\setlength{\tabcolsep}{4pt}
\renewcommand{\arraystretch}{1.05}
\begin{minipage}[t]{0.43\textwidth}
\vspace{0pt}
\begin{tabular}[t]{@{}p{0.24\linewidth}p{0.72\linewidth}@{}}
\toprule
\multicolumn{2}{@{}p{\linewidth}@{}}{\raggedright\textbf{Principles of Ability-Based Design\newline and How \mbox{\textit{AdaptType}} Addresses Them}} \\
\midrule
\raggedright \textbf{Principle} & \raggedright \textbf{How \mbox{\textit{AdaptType}} Addresses It} \tabularnewline
\midrule
\raggedright Ability & \raggedright
  The system is inherently centered around abilities by providing customization options and recommendations based on each user's gross and fine motor abilities. \tabularnewline \addlinespace[5pt]
\raggedright Accountability & \raggedright
  The system is adaptable by the user in its provision of a rich set of configuration options. \tabularnewline \addlinespace[5pt]
\raggedright Availability & \raggedright
  While this system is built on commodity hardware and software, future work should support deployment of the system on lower-cost devices than the Meta Quest Pro. \tabularnewline \addlinespace[5pt]
\raggedright Adaptability & \raggedright
  The system enables VR text entry methods that can be adapted to suit each user. \tabularnewline \addlinespace[5pt]
\raggedright Transparency & \raggedright
  The system provides visual and audio feedback on user actions, and always allows users to override system recommendations. \tabularnewline \addlinespace[5pt]
\raggedright Performance & \raggedright
  Future work could track a user's text entry performance in real time to suggest potential changes and improvements in a user's customizations. \tabularnewline \addlinespace[5pt]
\raggedright Context & \raggedright
  Future work could detect a user's task, activity, or situational context and make suggestions for adjustments accordingly. 
  \tabularnewline
\bottomrule
\end{tabular}
\end{minipage}%
\hfill%
\begin{minipage}[t]{0.54\textwidth}
\vspace{0pt}
\begin{tabular}[t]{@{}p{0.22\linewidth}p{0.75\linewidth}@{}}
\toprule
\multicolumn{2}{@{}p{\linewidth}@{}}{\raggedright\textbf{Ability Heuristics\newline and How \mbox{\textit{AdaptType}} Addresses Them}} \\
\midrule
\raggedright \textbf{Heuristic} & \raggedright \textbf{How \mbox{\textit{AdaptType}} Addresses It} \tabularnewline
\midrule
\raggedright Adaptability & \raggedright
  The system provides a highly customizable framework that enables users to configure input methods to their motor abilities. \tabularnewline \addlinespace[5pt]
\raggedright Equitable Experience & \raggedright
  Although user choices varied, the customized text entry method yielded consistent high performance and usability across participants who chose different methods. \tabularnewline \addlinespace[5pt]
\raggedright Flexible Task Completion & \raggedright
  Users can freely switch between input modalities, and auxiliary input pointers are provided. \tabularnewline \addlinespace[5pt]
\raggedright Efficiency and Effectiveness \\of User Action & \raggedright
  Users can enter text in VR with ease using the system, with high text entry performance and low task load. \tabularnewline \addlinespace[5pt]
\raggedright Multiple Modalities & \raggedright
  The system combines the gaze, hand and finger, head, and voice modalities to enable a multimodal text entry experience. \tabularnewline \addlinespace[5pt]
\raggedright Understandable Messages & \raggedright
  The system provides clear instructions for users and offers hints for conversational commands. \tabularnewline \addlinespace[5pt]
\raggedright Ease of Adoption & \raggedright
  The system runs on commodity hardware, while future work could expand its applicability to lower-cost devices. \tabularnewline \addlinespace[5pt]
\raggedright Ability Data Transparency & \raggedright
  Future work could improve the transparency of ability data, and inform users of how their ability data is being used to produce recommendations. \tabularnewline \addlinespace[5pt]
\raggedright Help, Support, \& Community & \raggedright
  Future work could offer support for troubleshooting and community feedback or support. \tabularnewline
\bottomrule
\end{tabular}
\end{minipage}
\vspace{0.3pc}
\caption{How \textit{AdaptType} addresses the seven principles of Ability-Based Design~\cite{wobbrock2018ability} (left) and the nine Ability Heuristics~\cite{mitchell2026ability} (right).}
\Description{Comparison of AdaptType with the seven principles of Ability-Based Design and nine Ability Heuristics. For each principle or heuristic, the table explains how AdaptType currently addresses it or identifies opportunities for future work. AdaptType particularly supports ability-centered customization, adaptability, user control, flexible task completion, multiple modalities, and understandable feedback, while availability, performance-aware adaptation, contextual adaptation, ability-data transparency, and community support remain areas for future development.}
\label{tab:abd-principles}
\vspace{-2pc}
\end{table*}

\subsection{Addressing Ability-Based Design Principles and ``Consequences'' in Accessibility}

The goal of \textit{AdaptType} is to accommodate a range of motor abilities, which aligns with the core idea of Ability-Based Design~\cite{wobbrock2011ability, wobbrock2018ability}. Thus, we examined our system against the seven principles of Ability-Based Design~\cite{wobbrock2018ability} (Table~\ref{tab:abd-principles}, left). \textit{AdaptType} addresses most principles through its highly customizable interface, while some principles remain open for future work, such as \textit{Performance} and \textit{Context}. Additionally, we examined our system against the nine Ability Heuristics (Table~\ref{tab:abd-principles}, right), which are used to conduct ability-based accessibility inspections~\cite{mitchell2026ability}. \textit{AdaptType} satisfies most heuristics, particularly through its highly flexible task flow and multimodal approach, while some heuristics point to future work improvements, such as \textit{Ability Data Transparency} and \textit{Help, Support, and Community}. 

Prior work has also emphasized that accessibility should be considered not only in terms of ``technical feasibility,'' but also in terms of the broader ``consequences'' of usage~\cite{mack2022chronically}. This perspective highlights that technologies should not simply make interaction possible for people with disabilities, but should also account for discomfort that may arise from their use. A technology should not be described as ``accessible'' if it causes significant fatigue, discomfort, or other negative consequences after use. In our study, some participants were technically able to ``use'' the state-of-the-art method based on hand pointing and pinch gestures. For example, P4 achieved seemingly reasonable text entry accuracy with \textit{Hand} (2.15\% uncorrected error rate), yet reported experiencing severe fatigue with their hand after. In contrast, \textit{AdaptType} mitigates the possible negative consequences of usage, and no participant reported having severe fatigue with the \textit{Custom-Iterated} technique despite it was the last technique evaluated. 

\subsection{Design Implications and Generalizability Beyond Virtual Reality}

Aside from its artifact contributions, this work also contributes empirical insights and design implications on personalized text entry for accessibility. One major co-design insight that informed our system was that participants were overwhelmed by the large number of customization options. Prior work found that while customization is often desirable, users may struggle to navigate the available options and may not be aware of which options work best for them~\cite{findlater2004comparison, roy2019automation, caraban201923, findlater2010beyond, stuerzlinger2006user}. Our finding aligns with prior work, and showed the effectiveness of ability-based recommendations in addressing this ``awareness'' gap by scaffolding customization using AI recommendations (Section~\ref{sec:finding-4}). 
Additionally, we found that iterative customization exhibited trends of improvement in text entry performance, task load, and usability (Section~\ref{sec:finding-5}). With highly configurable systems, users may experience steep learning curves with various modalities (``with practice I'll definitely get better,'' P9), and iteration through trial-and-error may improve the outcome of customization. This finding also aligns with prior work showing users' customization preferences evolve with usage~\cite{kong2025supporting, peracha2026personalized}. Together, our findings suggest that  personalized text entry interfaces should: (1) \textbf{\textit{offer ability-based recommendations to reduce the selection space and scaffold customization}}, and (2) \textbf{\textit{guide users to iteratively refine their customization choices over time}}. 

While our system focuses on text entry in VR, its end-user customization framework may generalize beyond virtual environments. We selected the VR scenario because the head-mounted display served as an integrated device that supported a wide range of sensing and interaction capabilities, including gaze tracking, hand and finger tracking, head tracking, voice input, and interface rendering. That said, similar functionality can be achieved outside VR using installed devices (e.g., the Tobii eye tracker and the OptiTrack motion capture system), enabling deployment without a headset or a fully virtual environment. Participants also expressed interest in using \textit{AdaptType} beyond VR contexts. For example, P7 noted that ``\textit{it'll be very cool to use this keyboard without the headset, so I could type with my eyes at home},'' highlighting the potential transferability of \textit{AdaptType}'s customization framework to broader multimodal text entry scenarios with personalized interaction techniques.

\subsection{Limitations and Future Work}

One limitation of this work is the small sample size (10 participants, 782 trials), which significantly limits the power of inferential analyses and generalizability of our findings. Due to the small sample size, we rely on descriptive statistics alongside inferential statistics to illustrate trends and showcase the capabilities of \textit{AdaptType}. We also discuss individual differences among participants, which aligns with the heterogeneity of the user population.

With regard to its implementation, \textit{AdaptType} relies on state-of-the-art gaze- and hand-tracking through the Meta Quest Pro and the OVREye, OVRHand, and OVRSkeleton libraries. Tracking accuracy and latency remain limiting factors, and future systems could benefit from higher-fidelity tracking techniques. A primary contribution of \textit{AdaptType} was its end-user customization framework rather than a novel gesture recognition technique; thus, our system used threshold-based heuristics with user calibration, which may be less robust than personalized gesture recognition models trained from a set of collected user gestures. Future work could involve more robust recognition techniques, for example, similar to the \$1 gesture recognizer~\cite{wobbrock2007gestures} or using transfer learning~\cite{pan2009survey}. Furthermore, our custom finger and hand/arm gesture detection relies on visible motion, which is inherently limited for users who retain little to no visible upper-body motor function. Future systems could incorporate additional sensing modalities such as electromyography (EMG) to detect muscle activations that do not produce observable motion. 

In our study, participants were allowed to customize numerous aspects of the system, from keyboard size and position to pointing and triggering modalities, utilizing a wide range of interaction techniques. Controlled comparisons accounting for all personalization factors can be hardly captured within a single experiment, and the effect of each specific factor on the performance differences remain unclear. Thus, our study should be regarded as an exploratory study showcasing the customization capability of \textit{AdaptType} and its benefits, rather than a comprehensive comparison of various input modalities or customization options. The \textit{Custom-Iterated} condition was also not counterbalanced like the other four conditions, as this condition required completion of the other four. Although we removed this condition from the cross-modality analyses, this limitation may result in biases that favored the \textit{iterated} customization when compared to the \textit{initial} customization. Additionally, only six participants completed the the \textit{iterated} customization, as extended VR headset interactions beyond two hours can cause severe fatigue and discomfort for participants with upper-body motor impairments. During our study, we observed a steep learning curve with the various techniques, and the two-hour study was relatively short for full familiarization. Given the limited session duration, our findings regarding iterative customization were only exploratory, and future work should include large-scale and longitudinal studies on how text entry behavior and customization preferences evolve over time. Furthermore, we included the \textit{editing} task in our study to showcase the benefits of multimodality in error correction, as natural error correction behavior can be hard to control and analyze in a lab study. That said, our editing task remains rather artificial, and future work should investigate natural error correction behavior with real-world tasks ``in the wild.''

%% file: 08-conclusion.tex
\section{Conclusion}

We have presented \textit{AdaptType,} an accessible and customizable multimodal VR text entry system that can be adapted to the diverse abilities and needs of people with upper-body motor impairments. By supporting a flexible combination of input modalities, offering a rich set of configuration settings, and providing ability-based recommendations to scaffold customization, our system allows users to create personalized VR text entry experiences that align with their abilities. Through a lab study with 10 participants with a range of upper-body motor disabilities, we found that customized multimodal techniques improved text entry speed compared to gaze- and hand-only techniques, and complemented voice input for error correction. Participants selected customization options that matched their upper-body motor abilities, achieved high text entry speed and accuracy with low task load. Participants found their customized techniques easy to use and learn, comfortable, causing little fatigue, and welcomed for future use. Our findings demonstrated the capability of \textit{AdaptType} to accommodate a wide range of motor abilities, and supporting ability-based accessible VR text entry through end-user customization.